\documentclass[journal,10pt]{IEEEtran}
\usepackage{algorithm}
\usepackage{algorithmic}

\usepackage{amssymb}
\usepackage{amsmath}
\usepackage{url}
\usepackage{xcolor}
\usepackage{cite,graphicx,amsmath,amssymb}
\usepackage{subfigure}
\usepackage{fancyhdr}
\usepackage{mdwmath}
\usepackage{mdwtab}
\usepackage{caption}
\usepackage{amsthm}
\usepackage{verbatim}

\newtheorem{theorem}{Theorem}

\newtheorem{lemma}{Lemma}

\newtheorem{corollary}{Corollary}

\newtheorem{proposition}{Proposition}
\usepackage{stfloats}

\makeatletter
\def\ScaleIfNeeded{%
	\ifdim\Gin@nat@width>\linewidth \linewidth \else \Gin@nat@width
	\fi } \makeatother

\begin{document}
	
	\title{Dual-Functional Artificial Noise (DFAN) Aided Robust Covert Communications in Integrated Sensing and Communications}

	\author{
		Runzhe~Tang,
		Long~Yang,~\IEEEmembership{Senior Member,~IEEE,}
		Lu~Lv,~\IEEEmembership{Member,~IEEE,}
		\\ Zheng~Zhang,~\IEEEmembership{Graduate Student Member,~IEEE,}
		Yuanwei~Liu,~\IEEEmembership{Fellow,~IEEE,} and
		Jian~Chen,~\IEEEmembership{Member,~IEEE}
		\thanks{Manuscript received Dec 26, 2023; revised March 10, 2024 and June 28, 2024; accepted August 21, 2024. This work was supported in part by the National Natural Science Foundation of China under Grant 62271368, in part by the Innovation Capability Support
			Program of Shaanxi under Grant 2024ZC-KJXX-080 and Grant 2024RS-CXTD-01, in part by the Key Research and Development Program of Shaanxi under
			Grant 2023-ZDLGY-50, in part by the Key Research and Development Program of Shaanxi under Grant 2023-YBGY-041, and in part by the Fundamental Research Funds for the Central Universities (Corresponding Author: Long Yang).}
		\thanks{R. Tang, L. Yang, L. Lv, Z. Zhang, and J. Chen are with the State Key
			Laboratory of Integrated Services Networks, Xidian University, Xi’an 710071,
			China (e-mail: tangrunzhe02@gmail.com; lyang@xidian.edu.cn; lulv@xidian.edu.cn; zzhang\_688@stu.xidian.edu.cn; jianchen@mail.xidian.edu.cn;).}
		\thanks{Y. Liu is with the Department of Electrical and Electronic Engineering (EEE) at The University of Hong Kong (HKU), Hong Kong, China (e-mail:
			yuanwei@hku.hk).}}

	\maketitle
	\vspace{-2cm}

	\begin{abstract}
		This paper investigates covert communications in an integrated sensing and communications system, where a dual-functional base station (called Alice) covertly transmits signals to a covert user (called Bob) while sensing multiple targets, with one of them acting as a potential watcher (called Willie) and maliciously eavesdropping on legitimate communications. To shelter the covert communications, Alice transmits additional dual-functional artificial noise (DFAN) with a varying power not only to create uncertainty at Willie’s signal reception to confuse Willie but also to sense the targets simultaneously. Based on this framework, the weighted sum of the sensing beampattern means square error (MSE) and cross correlation is minimized by jointly optimizing the covert communications and DFAN signals subject to the minimum covert rate requirement. The robust design considers both cases of imperfect Willie’s CSI (WCSI) and statistical WCSI. Under the worst-case assumption that Willie can adaptively adjust the detection threshold to achieve the best detection performance, the minimum detection error probability (DEP) at Willie is analytically derived in the closed-form expression. The formulated covertness constrained optimization problems are tackled by a feasibility-checking based difference-of-convex relaxation (DC) algorithm utilizing the S-procedure, Bernstein-type inequality, and the DC method. Simulation results validate the feasibility of the proposed scheme and demonstrate the covertness performance gains achieved by our proposed design over various benchmarks.
		
	\end{abstract}	

	\section{Introduction}
	The ubiquitous deployment of radar and communication (R$\& $C) systems leads to explosively growing demands for wireless resources (i.e., spectral and spatial resources). To fully exploit the potential of limited wireless resources as well as to certify the applications of simultaneous R$\& $C functions, a new paradigm denoted as integrated sensing and communications (ISAC) is proposed, which can provide R$\& $C functions on a single hardware platform with a single waveform \cite{isac1,isac2,isac3}. Due to the R$\& $C integration and coordination gains brought by ISAC technology, it has been envisioned as a key enabler for both
	next-generation wireless networks and radar systems \cite{isac4}, \cite{isac5}. Benefiting from the development of ISAC technology, the communication-based network design is shifted to ISAC networks, which enables communication-based networks to provide new high-accuracy sensing services such as unmanned aerial vehicles (UAVs) \cite{uav1}, \cite{uav2}, industrial Internet-of-things (IoT) \cite{iot} and intelligent traffic monitoring \cite{vehicular}.

	
	However, ISAC networks are confronted with severe security issues due to not only the broadcast characteristic of wireless communications but also the fact that sensed targets can potentially wiretap the confidential information-bearing signal transmitted to the legitimate receivers, which poses a new security concern \cite{pls1}. Although focusing power toward the target direction can facilitate sensing performance, the confidential information-bearing signal power for illuminating the target should be confined to prevent eavesdropping. Against this background, physical layer security (PLS) could be a viable approach to secure ISAC systems \cite{secure1}, \cite{secure2}. Different from conventional cryptographic techniques, which encrypt confidential data prior to transmission, PLS exploits the intrinsic randomness of noise and fading channels to degrade legitimate information leakage \cite{pls0}. There have been extensive works investigating
	the secrecy issue of ISAC systems in terms of the PLS \cite{pls2,pls3,pls4,pls5}.

	Compared with the PLS technology, covert communications aims to shelter the communication itself, which can achieve a higher level of security \cite{covert1}, \cite{covert2}. In certain circumstances, protecting the content of communications using existing PLS techniques is not sufficient as the communication itself is required to ensure a low probability of detection by adversaries, which motivates the studies of covert communications in ISAC networks \cite{covertisac1,covertisac2,covertisac3,covertisac4}. In \cite{covertisac1}, the authors proposed a covert beamforming design for ISAC systems. In \cite{covertisac2}, a robust transceiver design for a covert ISAC system is proposed. The authors in \cite{covertisac3} proposed an ISAC-aided covert transmission scheme, where a shared beamforming vector was considered for both sensing and communication and the warden Willie was equipped with multiple antennas. The authors in \cite{covertisac4} investigated the covert transmission in a general radar-communication cooperation system, where the radar cooperatively transmits sensing signals to track and simultaneously jam the aerial adversary target. Moreover, the aforementioned works all considered imperfect channel state information (CSI). Generally, covert communications are divided into two cases, i.e, infinite blocklength (the number of channel uses) and finite blocklength. When the blocklength is finite, the miss detection probability and the false alarm probability are complicated, thus Pinsker’s inequality can be used to obtain a lower bound of the detection error probability (DEP). Different from the aforementioned works studying covert communications in ISAC that based on finite blocklength hypothesis, we consider the case of infinite blocklength, which can be justified as follows. First, the assumption that Willie can obtain an infinite number of symbols establishes an upper bound on the number of symbols received by Willie, thus a upper bound of the DEP can be further invesigated. However, when the blocklength is finite, Pinsker’s inequality can only be used to obtain a lower bound of the DEP. Second, as indicated in \cite{covertisac5}, the number of symbols transmitted in each time slot is directly related to the bandwidth, thus the bandwidth can be properly selected to make the number of symbols transmitted in each time slot sufficiently large.
	
	For an arbitrary covert communications system, the availability of CSI at the base station (BS) (called Alice) is fundamental in covert communications to conceal the signal from being detected by a potential watcher (called Willie) \cite{covert3}. Considering a covert ISAC system working in the tracking mode \footnote{The sensing working mode comprises detection mode and tracking mode, which is selected based on the sensing requirements and the desired sensing beampattern is different. Generally, in the detection mode, due to the BS does not has prior information of targets, an isotropic beampattern is desired to reap the estimates of the target parameters, including the number of targets and their angles. On the other hand, in the tracking mode, the target parameters obtained in the detection mode can be harnessed to facilitate joint sensing and communication performance, i.e., the beampattern is designed to have prominent peaks in the target directions.}, the instantaneous CSI of the Alice-covert user (called Bob) link is available at the Alice as the CSI of Bob can be obtained via conventional channel acquisition techniques. However, the exact prior information of targets (the number of targets and their initial angle estimation) is usually unknown because the initial angle estimation of targets contains a certain degree of uncertainty \cite{covert4}. As illustrated above, it is necessary to consider imperfect WCSI in the covert ISAC system, which is consistent with the WCSI hypotheses in previous studies on covert ISAC systems \cite{covertisac1,covertisac2,covertisac3,covertisac4}. On the other hand, the worst-case scenario should be considered that only the statistical WCSI is available at Alice for robust covert ISAC system design. Due to Alice cannot obtain Willie’s instantaneous CSI, it is impossible for Alice to know Willie’s instantaneous error detection state, thus achieving covert communications is more challenging in this scenario. Nevertheless, there have been no prior works studying statistical WCSI in covert ISAC systems, which motivates the investigation into robust design of covert ISAC systems.
	
	To achieve covert communications and ensure a negligible probability of being detected by Willie, uncertainty should be created to confuse Willie (e.g. noise \cite{covert5}, channel uncertainty \cite{covert6}, full-duplex receiver \cite{covert7} and artificial noise (AN)\footnote{Generally, in conventional secure communications, AN acts as an interference signal to degrade the eavesdropper's received Signal-to-Noise Ratio (SNR). While in covert communications, AN is harnessed to conceal the covert communications signal. Specially, the warden's energy detection is confused by introducing random AN as an additional extrinsic uncertainty at the warden to achieve covert communications.} \cite{covert8}). In the ISAC system, the joint sensing and communication mechanism sheds light on the new secure design that the additional sensing function can serve as a support to facilitate the provision of security \cite{plsisac}, which raises the reflection on the potential interplay between ISAC and covert communications \cite{covertisac1,covertisac2,covertisac4}. To elaborate, in the ISAC system design, due to the degree-of-freedom (DoF) of conventional radar being limited by the number of transmit antennas, the dedicated sensing signal can be introduced to exploit the full DoF of radar function, especially in the cases where the number of users is smaller than the number of antennas that may lead to significant distortion of radar beampattern \cite{dedicated}. Specifically, in references \cite{covertisac1,covertisac2,covertisac4}, the radar beaming was harnessed to confuse Willie. However, the aforementioned works all considered the finite blocklength case that only a lower bound of the DEP can be derived. Hence, in this paper, the infinite blocklength case is investigated, where the power of the dedicated sensing signal changed randomly to achieve covert communications. Note that the dedicated sensing signal design in this paper is basically different from the dedicated radar signal design in references \cite{covertisac1,covertisac2,covertisac4}. To elaborate, in the infinite blocklength system design, uncertainties of the transmitted signals and the received additive white Gaussian noise (AWGN) will vanish, which necessitates introducing additional extrinsic uncertainties to achieve covert communications.

	Concerning the sensing performance metric, all the aforementioned studies of covert communications in ISAC networks adopted the sensing SNR (a general metric) to evaluate sensing performance. As a General Metric, the sensing SNR is applicable to various scenarios and the mathematical treatment is undemanding. However, they may not precisely and clearly characterize the practical sensing performance \cite{sensingmetric}. Hence in this paper, we harness the weighted sum of the beampattern matching mean squared error (MSE) and cross-correlation patterns (a classical radar metric) as the sensing performance metric.  Although adopting classical radar metrics to measure the performance of diverse sensing tasks also introduces more technical challenges compared to general metrics, utilizing classical radar metrics in covert ISAC networks can showcase the fundamental performance limits directly, which is beneficial for Alice to derive warden’s information and is also favorable for the covert ISAC system design.
	
	\subsection{Contributions and Organization}
	In this paper, we investigate infinite blocklength covert communications in an ISAC system, which consists of a dual-functional multi-antenna BS Alice, a single-antenna covert user Bob, and multiple sensing targets, where Alice transmits additional dual-functional artificial noise (DFAN) with a varying power to create uncertainty at Willie’s signal reception to shelter the covert communications and sense targets simultaneously. The main contributions of this work are summarized below:
	
	\begin{itemize}
		\item  We investigate an infinite blocklength covert ISAC system where sensing targets may act as a potential Willie. To shelter the covert communications, Alice transmits additional DFAN with varying power not only to confuse Willie but also to harness the DFAN to sense the targets simultaneously. 
		\item The robust covert ISAC design considers not only imperfect WCSI (i.e., bounded and Gaussian WCSI errors) but also statistical WCSI. Under both WCSI hypotheses, the worst-case scenario is considered that Willie can adaptively adjust the detection threshold to achieve the best detection performance. On this basis, the minimum DEP at Willie is derived in the analytical expressions. Moreover, in the statistical WCSI case, the closed-form expressions of the average minimum DEP are derived to further investigate the covertness of the system.
		\item The weighted sum of the sensing beampattern MSE and cross correlation is minimized by jointly optimizing the covert communications and DFAN signals subject to the minimum covert rate requirement, which is a more practical sensing performance metric compared to generic sensing metrics. Firstly, in the imperfect WCSI scenario, the formulated covert rate constrained and outage probability constrained optimization problems are tackled utilizing S-procedure, Bernstein-type inequality, and the difference-of-convex (DC) relaxation method. Next, in the statistical WCSI scenario, the closed-form expressions of the average minimum DEP are derived in the first place and then the formulated optimization problem is solved by the DC relaxation method. Generally, a feasibility-checking based DC algorithm design is proposed to solve the formulated optimization problems, which ensures the initial value of $\mathbf{W}_1$ in the feasible region of the problem and guarantees the convergence of the algorithm.
		
		\item Simulation results validate the feasibility of the proposed scheme. The results show that: On the one hand, to achieve covert communications in an ISAC system, DFAN is preferred rather than single-functional AN. On the other hand, when the covert rate or the sensing performance requirement is low, it is preferable to adopt the statistical WCSI hypothesis to improve the robustness of the covert ISAC design, while achieving comparable performance.
	\end{itemize}

	The remainder of this paper is organized as follows. Section
	II presents the system model. In Sections III and IV the covert communications performance analysis is illustrated firstly and the covert ISAC optimization problems are accordingly formulated and then solved under the imperfect and statistical WCSI scenario, respectively. Our algorithm design is presented in Section V. Section VI presents numerical results and a particular channel setting case is investigated in Section VII. Finally, Section VIII concludes this paper.
	
	\textit{Notations:} Boldface capital $\mathbf{X}$ and lower-case letter $\mathbf{x}$ denote matrix and vector, respectively. For any $N\times M$-dimensional matrix $\mathbf{X}\in\mathbb{C}^{N\times M}$, $\mathbf{X}^{T}$ and $\mathbf{X}^{H}$ denote the transpose and Hermitian conjugate operations. Similarly, $\text{rank}(\mathbf{X})$, $\text{Tr}(\mathbf{X})$, $\|\mathbf{X}\|$, $\|\mathbf{X}\|_{\text{F}}$ represent the rank value, trace value, spectral norm operation, and Frobenius norm operation. $\mathbf{X}\succeq\mathbf{0}$ denotes that $\mathbf{X}$ is a positive semidefinite matrix, while $\mathbf{x}\sim \mathcal{CN}(\mu,\mathbf{X})$ denotes that $\mathbf{x}$ is a circularly symmetric complex Gaussian (CSCG) vector with mean $\mu $ and covariance matrix $\mathbf{X}$. For a matrix $\mathbf{X}$, $\mathbf{X}^{-1}$ denotes the  inverse matrix operation. For any vector $\mathbf{x}$, $|x|$ and $\|\mathbf{x}\|$ denote the modulus of $x$ and the Euclidean norm of the vector $\mathbf{x}$, respectively. $\mathbb{E}($\textperiodcentered$)$ is the statistical expectation operation.

	\begin{figure} [t!]
		\centering
		\includegraphics[width= 2.3in, height=2.2in]{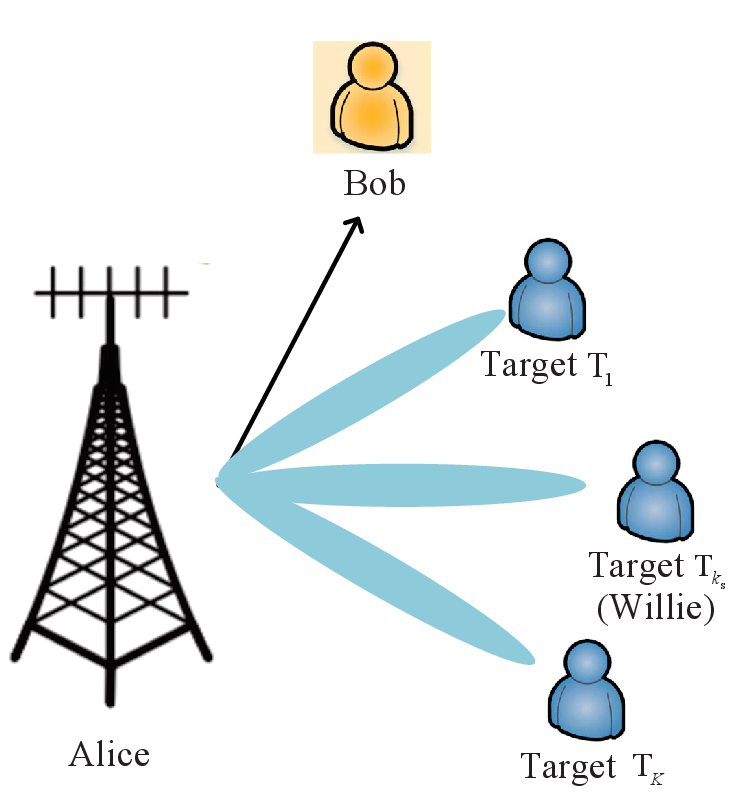}
		\caption{Covert communications in an ISAC system.
		}
		\label{system_model}
	\end{figure}
	
	\section{System Model}
	
	As shown in Fig. 1, we consider covert communications in an ISAC system, which consists of a dual-functional $N$-antenna BS called Alice, a single-antenna covert user called Bob, and $M$ sensing targets indexed by ${\cal M} = \{ T_{1},\cdots, T_{M}\} $ ($M \le N$). A challenging surveillance scenario is considered, where the sensing target  $T_{m_{\text{s}}}$ is assumed to be the potential warden called Willie, which aims to maliciously wiretap the communications between Alice and Bob. Note that the covert communications scenario considered in this work can be readily extended to the scenario with multiple sensing targets which are suspicious wardens, where AN can be transmitted in all the directions of suspicious wardens not only to achieve covert communications but also to sense these targets simultaneously. To achieve covert communications between Alice and Bob, Alice transmits additional DFAN to create uncertainty at Willie’s signal reception. Besides confusing Willie, the DFAN is also harnessed to sense $M$ targets simultaneously. The DFAN is independent of the covert communications signal and the total transmit power of the DFAN is denoted by $ {{P_{\rm A}}}$ and we assume that $ {{P_{\rm A}}}$ follows a uniform distribution within the range $\left[ {{P_{{\rm A},\min }},{P_{{\rm A},\max }}} \right]$, i.e.,
	\begin{align}	
		{f_{{P_{\rm A}}}}(x) = \frac{1}{{{P_{\rm A,\max }}}-{P_{\rm A,\min }}},{P_{\rm A,\min }} \le x \le {P_{\rm A,\max }},
	\end{align}
	where ${P_{\rm A,\max }}$ denotes the maximum transmit power budget for the DFAN. It is assumed that Willie knows the distribution of $ {{P_{\rm A}}}$. However, the instantaneous power of $ {{P_{\rm A}}}$ is not available at Willie. Due to the uncertainty introduced by the DFAN, Willie cannot tell whether the power fluctuation of its received signal is due to the variation of the ongoing covert communications or the variation of the DFAN. 
	In addition to assisting in the covert signal transmissions between Alice and Bob, i.e., guaranteeing a low probability of being detected by Willie, the DFAN is as well exploited to facilitate the sensing function.

	The channel coefficients from Alice to Bob and Willie are defined as ${\mathbf{h}}_{\rm b}\in {{\mathbb{C}}^{N\times 1}}$ and ${\mathbf{h}}_{\rm w}\in {{\mathbb{C}}^{N\times 1}}$, respectively. For small-scale fading, the Alice-Bob channel ${\mathbf{h}}_{\rm b}$ and the Alice-Willie channel ${\mathbf{h}}_{\rm w}$ are both assumed to be Rayleigh fading.The CSI availability is illustrated below. It is assumed that the covert ISAC system works in the tracking mode as the premise, where prior information of targets (the number of targets $M$ and their initial angle estimation ${\hat \theta _m}$) are known. We assume that Alice knows the instantaneous CSI of the Alice-Bob link. Moreover, the WCSI availability is divided into two scenarios, i.e., with imperfect WCSI and statistical WCSI. In the imperfect WCSI scenario, both bounded WCSI errors and Gaussian WCSI errors are investigated. Furthermore, in the statistical WCSI scenario, it is assumed that Willie can not acquire the instantaneous CSI of ${\mathbf{h}}_{\rm w}$, which is justified below. To facilitate achieving covert communications, it is assumed that Bob transmits pilot signals to Alice during channel estimation, thus Alice can estimate the downlink Alice-Bob channel and derive the instantaneous CSI of ${\mathbf{h}}_{\rm b}$ while the instantaneous CSI of ${\mathbf{h}}_{\rm w}$ is not available at Willie \cite{csi1}, \cite{csi2}. Moreover, the case that Willie can obtain the instantaneous CSI of ${\mathbf{h}}_{\rm w}$ is elaborated in section VII. The WCSI model is elaborated as follows.

	In general, imperfect WCSI refers to the case where the channel ${\mathbf{h}}_{\rm w}$ lies in some uncertainty set centered at the
	true channel from the perspective of Alice; Statistical WCSI means only the distribution of the channel ${\mathbf{h}}_{\rm w}$ is available at Alice. Moreover, within the regime of imperfect WCSI, bounded WCSI error model and Gaussian WCSI error model are two generic models. To elaborate, in the bounded WCSI error model, the CSI errors are assumed to lie within a bounded set. However, it is too strict and often leads
	to a very conservative design. On the other hand, the Gaussian WCSI error model is less restrictive and more practical by constraining occasional violation of the constraints.

	1) \textbf{Bounded WCSI errors:} In this scenario, Willie is also assumed to be a system user. Although there may be potential cooperation between Alice and Willie (e.g. Willie is an untrusted relay between the BS and users), only rough WCSI is available at Alice
	due to noisy measurements and/or rapidly changing channel conditions. Here, the bounded CSI error model is considered, i.e.,  
	\begin{align}	\label{0}
		{\mathbf{h}_{\rm w}} = {\hat {\mathbf{h}}_{\rm w}} + \Delta {\mathbf{h}}_{\rm w}, \left\| {\Delta {\mathbf{h}}_{\rm w}} \right\| \le \varepsilon_{\rm w},
	\end{align}
	where $\Delta {\mathbf{h}}_{\rm w}$ denotes the error vector, $\varepsilon_{\rm w}$ denotes the maximum threshold of the bounded CSI error, and ${\hat {\mathbf{h}}_{\rm w}}$ denotes the estimated CSI of Willie, i.e., sensing target $T_{m_{\text{s}}}$, at Alice.

	2) \textbf{Gaussian WCSI errors:} In this scenario, Willie is assumed to not be part of the system and the cooperation between Alice and Willie is limited. Alice needs to detect active transmission from Willie to acquire the WCSI. Moreover, Alice estimates the channel ${\mathbf{h}}_{\rm w}$ using the MMSE method, thus the CSI errors tend to follow
	Gaussian distribution. To elaborate, the WCSI is subject to Gaussian errors, i.e.,
	\begin{align}\label{120}
		{\mathbf{h}_{\rm w}} = {\hat {\mathbf{h}}_{\rm w}} + \Delta {\mathbf{h}}_{\rm w}={\hat {\mathbf{h}}_{\rm w}} + \boldsymbol{\gamma }_{\rm w}^{\frac{1}{2}}{\mathbf{e}_{\rm w}}, \Delta {\mathbf{h}}_{\rm w} \sim {\cal C}{\cal N}({\bold 0},{\boldsymbol{\gamma }_{\rm w}}),
	\end{align}
	where ${\boldsymbol{\gamma }_{\rm w}} = \boldsymbol{\gamma }_{\rm w}^{\frac{1}{2}}{\mathbf{e}_{\rm w}}$ is the covariance matrix of WCSI error and ${\mathbf{e}_{\rm w}}$ denotes the independent CSCG random vector following the distribution ${\mathbf{e}_{\rm w}} \sim {\cal C}{\cal N}({\bold 0},{\mathbf I})$. Moreover, the outage probability for covertness constraints is
	defined as ${\rho _c}$.

	3) \textbf{Statistical WCSI errors:}	To guarantee the robustness of the covert ISAC system, the worst scenario is considered that only the statistical WCSI is available at Alice. In this scenario, Willie is assumed to not be part of the system and keep silent. Alice needs to capture Willie’s leaked signals (the radiometer detector is adopted by Willie and involuntary signal leakage of radiometers is unavoidable) to identify some suspicious area and obtain the statistical WCSI. Some advanced detection methods, e.g., “Ghostbuster” introduced in \cite{ghost}, can be harnessed by Alice to acquire the corresponding statistical parameters. 
	Specifically, the channel from Alice to Willie is denoted as ${{\mathbf{h}}_{\rm w}}=\sqrt{{{l}_{\rm w}}}{{\mathbf{g}}_{\rm w}}$, where $\mathbf{g}_{\rm w}\sim \mathcal{C}\mathcal{N}({{{\bold 0}_{N \times 1}}},{{\mathbf{\Omega} _{\rm w}}})$ is the small-scale Rayleigh fading channel coefficient of the Alice-Willie link and the positive-semidefinite matrix ${\mathbf{\Omega} _{\rm w}}$ represents the spatial correlation matrix. The statistical characteristics of ${{\mathbf{h}}_{\rm w}}$ is firstly depicted and then the robust covert ISAC system design is further investigated according to it.

	\subsection{Transmission Scheme}
	Alice transmits DFAN to confuse Willie on the detection of the covert communications between Alice and Bob. The transmit DFAN signal at Alice is defined as ${{\boldsymbol x}_{\rm AN}}(k)=\sqrt {{P_{\rm A}}} {{\boldsymbol x}_{\rm A}}(k) $, where ${\left\| {{{\boldsymbol x}_{\rm A}}(k)} \right\|^2} = 1$ should be satisfied, with $k \in\{ 1,\cdots,K\} $ denotes the index of the signal symbol. 
	Let ${{\cal H}_0}$ denote the null hypothesis which indicates that Alice is not transmitting private data stream to Bob, while ${{\cal H}_1}$ denotes the alternate hypothesis which indicates an ongoing covert transmission
	from Alice to Bob. To elaborate, in $ {{\cal H}_0}$, the DFAN signal ${{\boldsymbol x}_{\rm AN}}(k)$ is transmitted to Willie, which aims at simultaneously sensing targets and confusing Willie. In $ {{\cal H}_1}$, in addition to the DFAN signal ${{\boldsymbol x}_{\rm AN}}(k)$, Alice transmits the covert communications signal ${\mathbf{w}}_1{s_{\rm b}(k)}$ to Bob. We use ${\mathbf{w}}_1\in {{\mathbb{C}}^{N\times 1}}$ to denote
	the corresponding transmit information beamforming vector in hypothesis ${{\cal H}_1}$. The information symbol ${s_{\rm b}(k)}$ is assumed to be statistically independent as well as with zero mean and unit power. Moreover, the DFAN signal ${{\boldsymbol x}_{\rm AN}}(k)$ is assumed to be independent with the information symbol and with the covariance matrix ${\mathbf{T}} = \mathbb{E}\left[ {{{\boldsymbol x}_{\rm AN}}(k){{\boldsymbol x}_{\rm AN}^H}(k)} \right]\succeq0$. Therefore, from Willie’s perspective, Alice’s transmitted signal is 
	given by 
	
	\begin{align}	\label{1}
		\boldsymbol{x}(k) = \left\{ \begin{array}{l}
			{{\boldsymbol x}_{\rm AN}}(k),\quad \quad \quad\quad\quad \thinspace\thinspace \thinspace \thinspace {{\cal H}_0},\\
			{\mathbf{ w}_1}{s_{\rm b}(k)} + {{\boldsymbol x}_{\rm AN}}(k),\quad   {{\cal H}_1}.
		\end{array} \right.
	\end{align}

	\subsection{Covert Communications}
	As illustrated in the last subsection, Alice transmits a DFAN signal ${{\boldsymbol x}_{\rm AN}}(k)$ and a covert
	signal ${s_{\rm b}(k)}$ to Bob. Here, Alice’s interference signal ${{\boldsymbol x}_{\rm AN}}(k)$ is exploited as a cover for Bob’s covert signal ${s_{\rm b}(k)}$. Note that due to Alice can obtain the instantaneous CSI of the downlink Alice-Bob channel, the channel capacity is precisely available to Alice. Hence the communication outage caused by DFAN can be prevented by keeping the transmission rate lower than the channel capacity during the variable rate transmission. Then, In $ {{\cal H}_1}$, the signal received at Bob is given by
	\begin{align}	\label{5}
		{y_b(k)} = \mathbf{h}_{\rm b}^H({\mathbf{w}_1}{s_{\rm b}(k)}+ {{\boldsymbol x}_{\rm AN}}(k))+ {n_{\rm b}(k)},
	\end{align}
	where ${n_{\rm b}} $ denotes the AWGN at Bob with zero
	mean and variance $\sigma _{\rm b}^2$. Thus the covert communication rate at Bob is
	given by
	\begin{align}	\label{7}
		{R_{\rm b}} = {\log _2}(1 + \frac{{{{\left| {\mathbf{h}_{\rm b}^H{\mathbf{w}_1}} \right|}^2}}}{{\mathbf{h}_{\rm b}^H{ {\mathbf{T}}} {\mathbf{h}_{\rm b}} + \sigma _{\rm b}^2}}),
	\end{align}
	where $\mathbf{h}_{\rm b}^H{ {\mathbf{T}}} {\mathbf{h}_{\rm b}}$ denotes the sensing interference power induced by DFAN. Note that in, $ {{\cal H}_0}$, the signal received at Bob is given by 
	\begin{align}	\label{1223}
		{y_b(k)} = \mathbf{h}_{\rm b}^H( {{\boldsymbol x}_{\rm AN}}(k))+ {n_{\rm b}(k)}.
	\end{align}
	Although in $ {{\cal H}_0}$ the received signal at Bob is treated as sensing noise which cannot be canceled due to the randomness of the DFAN. In this scenario, the DFAN can also be exploited to not only aid the secure communications between Alice and other legitimate users but also to sense the legitimate users simultaneously, which can possibly reduce the tracking overhead without the requirement for CSI feedback and the associated quantization and feedback errors.
	
	On the other hand, Willie tries to detect whether there exists covert communications between Alice and Bob or not by carrying out the
	Neyman-Pearson test based on his received signal sequence
	$	{y_w(k)}$ for $k \in\{ 1,\cdots,K\} $. Hence, as per \eqref{1}, the received signal at Willie can be expressed as
	\begin{align}	\label{8}
		{y_w(k)} = \left\{ \begin{array}{l}
			\mathbf{	h}_{\rm w}^H{{\boldsymbol x}_{\rm AN}}(k)+n_{\rm w}(k), \quad \quad \quad \quad \quad\thinspace\thinspace\thinspace   \thinspace \thinspace {{\cal H}_0},\\
			\mathbf{	h}_{\rm w}^H ({\mathbf{ w}_1}{s_{\rm b}(k)} + {{\boldsymbol x}_{\rm AN}}(k))+n_{\rm w}(k), \thinspace  \thinspace   {{\cal H}_1},
		\end{array} \right.
	\end{align}
	where ${n_{\rm w}} $ denotes the AWGN at Willie with zero
	mean and variance $\sigma _{\rm w}^2$.

	Based on the two hypothesis, Willie is assumed to adopt a radiometer for the binary detection. Note that the optimal test for Willie	to minimize the detection error probability is the likelihood
	ratio test on the grounds of the Neyman-Pearson criterion. However, the instantaneous WCSI is not available at Alice, which makes it difficult to directly analyze the detection performance at Willie. Using the average received power at Willie (i.e., $\varsigma = \frac{1}{K}\sum\nolimits_{k = 1}^K {{{\left| {{y_w}\left[ k \right]} \right|}^2}} $) as the test statistic, the decision rule is
	given by
	\begin{align}
		{\varsigma}\underset{{{D}_{0}}}{\overset{{{D}_{1}}}{\mathop{\gtrless }}}\,\Gamma,
	\end{align}
	where $\Gamma >0$ is Willie’s detection threshold,
	${D}_{1}$ and ${D}_{0}$ are the binary decisions in favor of ${{\cal H}_1}$ and ${{\cal H}_0}$, respectively. The case of infinite blocklength is considered, i.e., $K \to  + \infty $, where $\mathop {\lim }\limits_{n \to \infty } {{\chi _{2{\rm{n}}}^2} \mathord{\left/
			{\vphantom {{\chi _{2{\rm{n}}}^2} n}} \right.
			\kern-\nulldelimiterspace} n} = 1$ holds according to the Strong Law of Large
	Numbers. Thus the average received power at Willie can be obtained as
	
	\begin{align}
		\varsigma  \!=\! \left\{ {\begin{array}{*{20}{l}}
				{ {{P_{\rm A}}}{\left|\mathbf{	h}_{\rm w}^H{{\boldsymbol x}_{\rm A}}\right|}^2+ \sigma _{\rm w}^2, \thinspace\thinspace  \thinspace \quad \quad \quad \quad\quad  {{{\cal H}_0}},} \\
				{{{\left|\mathbf{	h}_{\rm w}^H{\mathbf{ w}_1} \right|}^2} +{{P_{\rm A}}}{\left|\mathbf{	h}_{\rm w}^H{{\boldsymbol x}_{\rm A}}\right|}^2+ \sigma _{\rm w}^2,\enspace{{{\cal H}_1}},}
		\end{array}} \right.
	\end{align}
	where the prior probabilities of hypotheses  ${{\cal H}_0}$
	and ${{\cal H}_1}$ are assumed to be equal for simplicity.
	
	Then, the detection error probability at Willie, $\xi $, is defined as 
	\begin{align}\label{171}
		\xi  \buildrel \Delta \over = {P_{\rm FA}} + {P_{\rm MD}}, 
	\end{align}
	where ${P_{\rm FA}} = P({D_1}|{{\cal H}_0})$ denotes the false alarm probability, ${P_{\rm MD}} = P({D_0}|{{\cal H}_1})$ denotes the miss detection probability, and $0 \le \xi  \le 1$. To be specific, $\xi  = 0$ implies that Willie can perfectly detect the covert signal without error, while $\xi  = 1$ implies that Willie cannot make a correct detection at the time, i.e., a blind guess.

	Based on the decision rule, we can derive the analytical expressions of ${P_{\rm FA}}$ and ${P_{\rm MD}}$ under three typical WCSI availability hypotheses which are elaborated before. The detailed analysis of Willie’s DEP will be elucidated in section \uppercase\expandafter{\romannumeral3} and section \uppercase\expandafter{\romannumeral4}. In addition, ${{\xi}^ * }\ge1-\epsilon$ is generally adopted as the covertness constraint in covert communications, where $\epsilon$ is a small value to
	determine the required covertness level and ${{\xi}^ * }$ denotes the minimum DEP.

	\subsection{Radar Sensing}
	To facilitate sensing design, we define the beampattern gain $P(\theta )$ as the transmit signal power distribution at sensing angle $\theta  \in \left[ { - \frac{\pi }{2},\frac{\pi }{2}} \right]$. $P(\theta )$ in ${{\cal H}_1}$ is given by
	
	\begin{align}	\label{15}
		P (\theta ) = \mathbb{E}({\left| {{\mathbf{a}^H}(\theta )({\mathbf{w}_1}{s_{\rm b}} + {\boldsymbol x}_{\rm AN})} \right|^2}) = {\mathbf{a}^H}(\theta )(\mathbf{W}_1+\mathbf{T})\mathbf{a}(\theta) ,
	\end{align}
	where $\mathbf{a}(\theta )$ denotes the steering vector given by $\mathbf{a}(\theta ) = {[1,{e^{j2\pi \frac{d}{\lambda }\sin (\theta )}},...,{e^{j2\pi (N - 1)\frac{d}{\lambda }\sin (\theta )}}]^T}$, $d$ denotes the antenna
	spacing and $\lambda $ denotes the carrier wavelength. Note that $\mathbf{T}$ denotes the DFAN signal covariance matrix.

	We assume that the sensing system works in the tracking mode and has prior information on targets (the number of targets $M$ and their initial angle estimation ${\hat \theta _m}$) are known. Thus the beampattern is expected to have the dominant peaks in the target directions. Given the estimated angles of $M$ sensing targets, the desired
	beampattern can be defined as a square waveform at the target directions given by
	
	\begin{align}	\label{17}
		{{\cal P}^ * }(\theta )= \left\{ \begin{array}{l}
			1,\quad \left| {\theta  - {\hat \theta _m}} \right| \le \frac{{\Delta \theta }}{2},\\
			0, \quad \rm {otherwise},
		\end{array} \right.
	\end{align}
	where ${\Delta \theta }$ is the desired beam width. Let $\{ {\theta _s}\} _{s = 1}^S$ denote the $S$ sample angles covering the detector’s angular range $\theta  \in \left[ { - \frac{\pi }{2},\frac{\pi }{2}} \right]$.
	Thus in ${{\cal H}_1}$, the MSE between the
	obtained sensing beampattern and the desired sensing beampattern, which is denoted as $F(\eta,\mathbf{W}_1,\mathbf{T})$, can be given by
	\begin{align}	\label{18}
		F(\eta_1,\mathbf{W}_1,\mathbf{T})=\frac{1}{S}	\sum\limits_{s = 1}^S {{{\left| {\eta_1 {{\cal P}^ * }({\theta _s}) - {\mathbf{a}^H}({\theta _s})(\mathbf{W}_1+\mathbf{T})\mathbf{a}({\theta _s})} \right|}^2}},
	\end{align}
	where $\eta$ denotes the scaling vector.
	
	To evaluate the radar sensing performance thoroughly, besides the sensing beampattern MSE, cross correlation among different transmit signal directions should also be considered. To elaborate, the cross correlation
	is given as
	\begin{align}
		{ L(\mathbf{W}_1,\mathbf{T})}=\frac{2}{{{M^2} - M}}\sum\limits_{p = 1}^{M } {\sum\limits_{q = p + 1}^M {{{\left| {\mathbf{a}^H}(\hat \theta_p )(\mathbf{W}_1+\mathbf{T})\mathbf{a}(\hat \theta_q ) \right|}^2}} } . 
	\end{align}
	Therefore, the weighted sum of the sensing beampattern MSE and cross correlation is considered as the sensing performance metric, which is given as
	\begin{align}
		L(\eta_1,\mathbf{W}_1,\mathbf{T})=F(\eta,\mathbf{W}_1,\mathbf{T})+w_c{\rm L(\mathbf{W}_1,\mathbf{T})},
	\end{align}
	where $w_c$\footnote{According to reference \cite{covert4}, in Fig. 3(a) all cross-correlation coefficients are approximately zero for $w_c=1$. While in Fig. 3(b) the cross-correlation behavior of the designs obtained with $w_c=1$ is much better than that of the designs obtained with $w_c=0$. Hence in this paper, we directly set the factor $w_c$ as $w_c=1$ and focus on the covert ISAC system design.} is a pre-determined weighting factor. Note that in ${{\cal H}_0}$ the sensing performance metric is similar to that elaborated in ${{\cal H}_1}$, the DFAN can be harnessed to sense targets while confusing Willie. In this work, we focus on the sensing beampattern design in ${{\cal H}_1}$. To elaborate, by optimizing $\mathbf{T}$ and ${\mathbf{w}_{ 1}}$, the weighted sum of the sensing beampattern MSE and cross correlation can be minimized and thus the optimal sensing performance can be derived under covert rate constraints.

	\section{Covert Communications under Imperfect WCSI}
	In this section, firstly, we delve into the worst-case covert communications performance in the considered covert ISAC system under the imperfect WCSI scenario (including bounded WCSI errors and Gaussian WCSI errors). To elaborate, Willie’s DEP is analyzed in the first place. As the worst case is considered that Willie can adopt the optimal detection threshold ${\Gamma ^ * }$ to minimize DEP to achieve the best detection performance, the optimal detection threshold, ${\Gamma ^ * }$, for Willie can be further derived. Thus the minimum DEP at Willie is derived in the analytical expressions. Then the covert communication rate constrained beampattern optimization problems are formulated under bounded WCSI errors and Gaussian WCSI errors. The formulated problems are non-convex and thus difficult to solve, which are tackled by leveraging the S-procedure, the Bernstein-type inequality (BTI) and the difference-of-convex (DC) relaxation method.

		\subsection{Covert Communications under Bounded WCSI Errors}
		Firstly, the covert communications performance is analyzed. To elaborate, based on the decision rule that $
		{\varsigma}\underset{{{D}_{0}}}{\overset{{{D}_{1}}}{\mathop{\gtrless }}}\,\Gamma
		$, the analytical expressions of $P_{\rm FA}$ and $P_{\rm MD}$ can be respectively given by
		\begin{align}
			{P_{\rm FA}} &= {\rm Pr}({P_ {\rm A}}{\rho _1} + \sigma _{\rm w}^2 > \Gamma ) \nonumber\\
			&= \left\{ \begin{array}{l}
				1,\quad\thinspace\thinspace \quad\quad\quad\quad\quad\quad\quad\quad\quad\thinspace\thinspace\thinspace\thinspace \Gamma<\sigma _{\rm w}^2,\\
				1 - \frac{{\Gamma  - \sigma _{\rm w}^2 - {P_{\rm A,\min }}{\rho _1}}}{{({P_{\rm A,\max }} - {P_{\rm A,\min }}){\rho _1}}},\sigma _{\rm w}^2 < \Gamma<{\Delta _1},\\
				0,\quad\quad\quad\quad\quad\quad\quad\quad\quad\quad\quad\Gamma>{\Delta _1},
			\end{array} \right.
		\end{align}
		
		\begin{align}
			{P_{\rm MD}} &= \Pr ({P_{\rm A }}{\rho _1} + {\rho _2} + \sigma _{\rm w}^2 < \Gamma )\nonumber\\
			&= \left\{ \begin{array}{l}
				0, \quad\quad\quad\quad\quad\quad\quad\quad\quad\thinspace\thinspace\thinspace\thinspace\Gamma<{\Delta _2}, \\
				\frac{{\Gamma  - {\rho _2} - \sigma _{\rm w}^2 - {P_{\rm A,\min }}{\rho _1}}}{{({P_{\rm A,\max }} - {P_{\rm A,\min }}){\rho _1}}},{\Delta _2} < \Gamma  <{\Delta _3},\\
				1,\quad\quad\quad\quad\quad\quad\quad\quad\quad\thinspace\thinspace\thinspace\thinspace\Gamma  >{\Delta _3},
			\end{array} \right.
		\end{align}

		where ${\rho _1} = {\left| {\mathbf{h}_{\rm w}^H{{\boldsymbol x}_{\rm A}}} \right|^2}, {\Delta _1}={P_{\rm A,\max }}{\rho _1} + \sigma _{\rm w}^2$ and ${\rho _2} = {\left| {\mathbf{h}_{\rm w}^H{\mathbf{w}_1}} \right|^2},  {\Delta _2}= {\rho _2} + \sigma _{\rm w}^2, {\Delta _3}={P_{\rm A,\max }}{\rho _1} + {\rho _2} + \sigma _{\rm w}^2$.
		
		As per (\ref{171}), the DEP at Willie can be derived. When ${\Delta _1} < {\Delta _2}$, the minimum value of $\xi$ is $\xi=0$. In this case it is impossible to achieve covert communications. When ${\Delta _1} > {\Delta _2}$, the DEP at Willie
		is given by
		\begin{align}
			\xi  = \left\{ \begin{array}{l}
				1,\quad\quad\quad\quad\quad\quad\quad\quad\quad\thinspace\thinspace \thinspace\thinspace\Gamma  < {\Delta _2},\\
				1 + \frac{{ - {\rho _2}}}{{({P_{\max }} - {P_{\min }}){\rho _1}}},{\Delta _2} < \Gamma  < {\Delta _1},\\
				\frac{{\Gamma  - {\rho _2} - \sigma _{\rm w}^2 - {P_{\rm A,\min }}{\rho _1}}}{{({P_{\rm A,\max }} - {P_{\rm A,\min }}){\rho _1}}},{\Delta _1} < \Gamma  < {\Delta _3},\\
				1,\quad\quad\quad\quad\quad\quad\quad\quad\quad\thinspace\thinspace \thinspace\thinspace\Gamma  > {\Delta _3}.
			\end{array} \right.
		\end{align}
		In the range $\left[ {{\Delta _1},{\Delta _3}} \right]$, $\xi$ is a monotonically increasing function of $\Gamma$. Thus, the minimum DEP lies in the range $\left[ {{\Delta _2},{\Delta _1}} \right]$, i.e., 
		\begin{align}
			{\xi ^ * } = 1 + \frac{{ - {\rho _2}}}{{({P_{\rm A,\max }} - {P_{\rm A,\min }}){\rho _1}}},
		\end{align}
		Note that ${\Delta _1} > {\Delta _2}$ should be satisfied as a prerequisite to achieving covert communications. Considering the covertness constraint ${{\xi}^ * }\ge1-\epsilon$ and ${\Delta _1} > {\Delta _2}$, we can respectively derive
		that 
		\begin{align}\label{172}
			\mathbf{h}_{\rm w}^H({\mathbf{W}_1} - \epsilon ({P_{\rm A,\max }} - {P_{\rm A,\min }})\mathbf{T}/P_{\rm A})\mathbf{h}_{\rm w} \le 0,
		\end{align}
		\begin{align}\label{173}
			\mathbf{h}_{\rm w}^H({\mathbf{W}_1} - {P_{\rm A,\max }}\mathbf{T}/P_{\rm A}){\mathbf{h}_{\rm w}} \le 0.
		\end{align}
		Notice that due to $\epsilon  \in \left[ {0,1} \right]$, the inequality $\epsilon ({P_{\rm A,\max }} - {P_{\rm A,\min }})\le {P_{\rm A,\max }}$ always holds, hence the covertness constraint (\ref{173}) is omitted.

		Our objective is to jointly optimize the covert communications
		signal and the DFAN signal such that the weighted sum of the sensing beampattern MSE and cross correlation is minimized in $ {{\cal H}_1}$,
		subject to the covert communications requirements under bounded WCSI errors. Based on the covertness analyses,
		the covert communication rate constrained beampattern optimization problem is formulated as 
		
		\begin{subequations}\label{190}
			\begin{align}\text {(P1.1)}:
				&\mathop {\rm min}\limits_{\eta_1,\mathbf{W}_1,\mathbf{T}}  	L(\eta_1,\mathbf{W}_1,\mathbf{T})
				\label{190:sub1}
				\\ &s.t.\thinspace\thinspace\thinspace\thinspace(21),\label{190:sub2}\\
				&\quad\quad\thinspace{R_{\rm b}} \ge {R_{\min }},\label{190:sub3}
				\\&\quad\quad\thinspace{\rm Tr}(\mathbf{W}_1)\le {P_{\rm t}}-{P_{\rm A,\max }},\label{190:sub5}
				\\
				&\quad\quad	\thinspace\thinspace{\rm rank}({\mathbf{W}_{\rm 1}}) = 1,\label{190:sub8}
			\end{align}	
		\end{subequations}
		where ${P_{\rm t}}$ denotes the total
		transmit power budget and ${R_{\min }}$ denotes the covert communication rate threshold. \eqref{190:sub2} is the covertness constraint in the scenario with bounded WCSI errors. Notice that problem $\text {(P1.1)}$ is non-convex due to the non-convex covertness constraint \eqref{190:sub2} and the covert communication rate constraint \eqref{190:sub3}, and the quadratic form of the beamformers makes the constraint \eqref{190:sub5} non-convex. Hence problem $\text {(P1.1)}$ is difficult to solve.

		To solve the formulated problem $\text {(P1.1)}$, firstly, the derived analytical covertness constraint (\ref{172}) is tackled by leveraging the S-procedure. To elaborate, according to the considered bounded WCSI model given in (\ref{0}), we can derive that
		$\Delta {\mathbf{h}}_{\rm w}^H\Delta {\mathbf{h}}_{\rm w} \le \varepsilon_{\rm w}^2 $. Then the S-procedure is introduced to convexify the covertness constraint (\ref{172}), which is given in the following lemma.
		\begin{lemma}
			(S-procedure): Suppose that ${f_i}(\mathbf{x}) = {\mathbf{x}^H}{\mathbf{A}_i}\mathbf{x} + 2{\mathop{\rm Re}\nolimits} \{ \mathbf{b}_i^H\mathbf{x}\}  + {c_i},i = 1,2,$ where ${\mathbf{x}}\in {{\mathbb{C}}^{N\times 1}}$,  ${\mathbf{A}_i}\in {{\mathbb{C}}^{N\times N}}$, ${\mathbf{x}}\in {{\mathbb{C}}^{N\times 1}}$, $\mathbf{b}_i\in {{\mathbb{C}}^{N\times 1}}$ and ${c_i} \in \mathbb{R}$. The condition ${f_1} \le 0 \Rightarrow {f_2} \le 0$ holds if and only if there exists a variable $\lambda $ such that
			\begin{align}
				\lambda \left[ {\begin{array}{*{20}{c}}
						{{\mathbf{A}_1}}&{{\mathbf{b}_1}}\\
						{\mathbf{b}_1^H}&{{c_1}}
				\end{array}} \right] - \left[ {\begin{array}{*{20}{c}}
						{{\mathbf{A}_2}}&{{\mathbf{b}_2}}\\
						{\mathbf{b}_2^H}&{{c_2}}
				\end{array}} \right]  \succeq {\bold 0}.
			\end{align}
		\end{lemma}
		According to
		Lemma 1, covertness constraints (\ref{172}) is equivalent to
		\begin{align}\label{bounded}
			\left[ {\begin{array}{*{20}{c}}
					{\lambda_1 \mathbf{I} - \mathbf{S}_1}&{ - \mathbf{S}_1{\hat {\mathbf{h}}_{\rm w}}}\\
					{ - {\hat {\mathbf{h}}_{\rm w}}^H\mathbf{S}_1}&{ - \lambda \varepsilon _{\rm w}^2 - {\hat {\mathbf{h}}_{\rm w}}^H\mathbf{S}_1{{\hat {\mathbf{h}}_{\rm w}}}}
			\end{array}} \right]\succeq {\bold 0},
		\end{align}

		where $\mathbf{S}_1={\mathbf{W}_1} - \epsilon ({P_{\rm A,\max }} - {P_{\rm A,\min }})\mathbf{T}/P_{\rm A}$ and $\lambda_1\ge0$ is the new auxiliary optimization variable.

		Next, to guarantee the rank-one property of $\mathbf{W}_{\rm 1}$, we adopt the DC relaxation method to extract the rank-one solution from the high-rank matrix \cite{dc}. In particular, \eqref{190:sub8} can reformulated as
		\begin{align}\label{12}
			\Re ({\rm Tr}(\mathbf{W}_{\rm 1}^H(\mathbf{I} - {\mathbf{w}_{ l}}\mathbf{w}_{ l}^H))) \le {\varrho _{\mathbf{w}_l}},
		\end{align}
		where ${\mathbf{w}_{l}} \in {{\mathbb{C}}^{N \times 1}}$ denotes the leading eigenvector of $\mathbf{W}_{\rm 1}$ derived in the previous iteration, and  $ {\varrho _{\rm w_1}} \to 0$ is the penalty factor. Thus the reformulated problem can be given as
		\begin{subequations}\label{1911}
			\begin{align}\text {(P1.2)}:
				&\mathop {\rm min}\limits_{\eta_1,\mathbf{W}_1,\mathbf{T}}  	L(\eta_1,\mathbf{W}_1,\mathbf{T})
				\label{1911:sub1}
				\\ &s.t.\thinspace\thinspace\thinspace\thinspace(25),(26),\label{1911:sub2}\\
				&\quad\quad\thinspace{\mathbf{h}_{\rm b}^H{\mathbf{W}_1}\mathbf{h}_{\rm b}}\ge{(2^{{R_{\min }} }- 1)}{{\mathbf{h}_{\rm b}^H{ {\mathbf{T}}} {\mathbf{h}_{\rm b}} + \sigma _{\rm b}^2}},\label{1911:sub3}
				\\&\quad\quad\thinspace{\rm Tr}(\mathbf{W}_1)\le {P_{\rm t}}-{P_{\rm A,\max }},\label{1911:sub5}
			\end{align}	
		\end{subequations}
		Problem $\text {(P1.2)}$ is a convex
		quadratic semidefinite programing (QSDP) problem. By utilizing off-the-shelf convex programming
		numerical solvers such as CVX \cite{cvx}, $\text {(P1.2)}$ can
		be optimally tackled.

		\subsection{Covert Communications under Gaussian WCSI Errors}
		Firstly, the covert communications performance is analyzed. The DEP at Willie can be derived similarly to the mathematical manipulations in the bounded WCSI errors scenario, which is thus omitted here. In the Gaussian WCSI errors scenario, the outage probability should be considered. Thus based on the covertness constraint (\ref{172}), the outage probability constraint for the covertness of the covert ISAC system can be given as 
		\begin{align}\label{121}
			&\Pr ( 	\mathbf{h}_{\rm w}^H\mathbf{S}_1\mathbf{h}_{\rm w}  \le 0) \ge 1 - {\rho _c},
		\end{align}
		where $\mathbf{S}_1={\mathbf{W}_1} - \epsilon ({P_{\rm A,\max }} - {P_{\rm A,\min }})\mathbf{T}/P_{\rm A}$.

		To tackle the the outage probability constraints (\ref{121}), the BTI is harnessed \cite{bti}, which is given in the following lemma
		\begin{lemma}
			(BTI): For any $\mathbf{A} \in {\mathbb{C}^{N \times N}},\mathbf{b} \in {\mathbb{C}^{N \times 1}},c \in \mathbb{R}, \mathbf{x}\sim {\cal C}{\cal N}(0,{\mathbf I})$ and $\rho  \in \left[ {0,1} \right]$, if there exist x and y, such that
			\begin{align}
				{\rm Tr}(\mathbf{A}) - \sqrt {2\ln (\frac{1}{\rho })}x  + \ln(\rho )y + c \ge 0,\\
				\sqrt {\left\|\mathbf{A} \right\|_F^2 + 2{{\left\| \mathbf{b} \right\|}^2}}  \le x,\\
				y\mathbf{I} + \mathbf{A} \succeq {\bold 0},y \ge 0,
			\end{align}
			the following inequality holds true
			\begin{align}
				{\rm Pr}({\mathbf{x}^H}{\mathbf{A}}\mathbf{x} + 2{\mathop{\rm Re}\nolimits} \{ \mathbf{x}^H\mathbf{b}\}  + {c} \ge 0)\ge 1-\rho.
			\end{align}
		\end{lemma}

		Recall that $	{\mathbf{h}_{\rm w}} ={\hat {\mathbf{h}}_{\rm w}} + \boldsymbol{\gamma }_{\rm w}^{\frac{1}{2}}{\mathbf{e}_{\rm w}}$. As per Lemma 2, (\ref{121}) can be equivalently transformed to the following inequalities
		\begin{align}
			{\rm Tr}(\mathbf{A}_{\rm w}) - \sqrt {2\ln (\frac{1}{\rho_c })}x  + \ln(\rho_c )y + c_{\rm w} \ge 0,\label{Gaussian1}\\
			\sqrt {\left\|\mathbf{A}_{\rm w} \right\|_F^2 + 2{{\left\| \mathbf{b}_{\rm w} \right\|}^2}}  \le x,\label{Gaussian2}\\
			y\mathbf{I} + \mathbf{A}_{\rm w} \succeq {\bold 0},y \ge 0,\label{Gaussian3}
		\end{align}
		where $\mathbf{A}_{\rm w}=\boldsymbol{\gamma }_{\rm w}^{\frac{1}{2}}(-\mathbf{S}_1)\boldsymbol{\gamma }_{\rm w}^{\frac{1}{2}}$, $c_{\rm w}={\hat {\mathbf{h}}_{\rm w}^{H}}(-\mathbf{S}_1){\hat {\mathbf{h}}_{\rm w}}$ and  $\mathbf{b}_{\rm w}=\boldsymbol{\gamma }_{\rm w}^{\frac{1}{2}}(-\mathbf{S}_1){\hat {\mathbf{h}}_{\rm w}}$.
		
		Similar to the bounded WCSI errors case, the non-convex rank-one constraint of $\mathbf{W}_{\rm 1}$ is also tackled by the DC relaxation method. Thus based on the covertness analyses,
		the covert communication rate constrained beampattern optimization problem is formulated as

		\begin{subequations}\label{191}
			\begin{align}\text {(P2)}:
				&\mathop {\rm min}\limits_{\eta_1,\mathbf{W}_1,\mathbf{T}}   	L(\eta_1,\mathbf{W}_1,\mathbf{T})
				\\ &s.t.\thinspace\thinspace\thinspace\thinspace(\ref{Gaussian1}),(\ref{Gaussian2}),(\ref{Gaussian3}),\label{191:sub2}\\
				&\quad\quad\thinspace{\mathbf{h}_{\rm b}^H{\mathbf{W}_1}\mathbf{h}_{\rm b}}\ge{(2^{{R_{\min }} }- 1)}{{\mathbf{h}_{\rm b}^H{ {\mathbf{T}}} {\mathbf{h}_{\rm b}} + \sigma _{\rm b}^2}},\label{191:sub3}
				\\&\quad\quad\thinspace{\rm Tr}(\mathbf{W}_1)\le {P_{\rm t}}-{P_{\rm A,\max }},\label{191:sub5}
				\\&\thinspace\thinspace\thinspace\thinspace\thinspace\thinspace\thinspace\thinspace\thinspace\thinspace\thinspace\thinspace(26),
			\end{align}	
		\end{subequations}
		where \eqref{191:sub2} denotes the outage-constrained constraints in the scenario with Gaussian WCSI errors, where the outage probability should be confined to a certain threshold $\rho_c$. 
		Problem $\text {(P2)}$ is a convex QSDP problem, which can be optimally tackled.

		\section{Covert Communications under Statistical WCSI}
		In this section, firstly, we dig into the worst-case covert communications performance in the considered covert ISAC system under the statistical WCSI scenario. To elaborate, Willie’s DEP is first analyzed 
		and the optimal detection threshold, ${\Gamma ^ * }$, for Willie is further derived considering the worst case that Willie can adopt the optimal detection threshold ${\Gamma ^ * }$ to minimize DEP to achieve the best detection performance. After the minimum DEP at Willie is derived in the analytical expressions, the
		closed-form expressions of the average minimum DEP are further derived to investigate the covertness of the system. Then the covert communication rate constrained beampattern optimization problem is formulated. The formulated problem is non-convex and thus difficult to solve, which is tackled by leveraging the DC relaxation method.
		\subsection{Covert Communications under statistical WCSI Errors}
		Firstly, the covert communications performance is thoroughly analyzed. We first derive the analytical expressions for
		${P_{\rm FA}}$ and ${P_{\rm MD}}$ in closed form. By analyzing the derived analytical expression of DEP, the optimal detection threshold
		${\Gamma ^ * }$ and the minimum DEP ${\xi ^ * }$ are obtained. Note that Willie cannot acquire the instantaneous CSI of ${\mathbf{h}}_{\rm w}$ and the case that Willie can acquire the instantaneous CSI of ${\mathbf{h}}_{\rm w}$ is depicted in section VII. Considering the fact that the
		instantaneous CSI of channel ${{\mathbf{h}}_{\rm w}}$ is not available at Alice, we adopt the
		minimum DEP averaging over variable ${t _{\rm A}}$ as the covertness metric of the considered covert ISAC system. To elaborate, the analytical expressions for
		${P_{\rm FA}}$ and ${P_{\rm MD}}$ can be respectively given as (\ref{23106}) and  (\ref{23107}), at the bottom of the next page, respectively.
		\begin{align}\label{23106}
			{P_{\rm FA}} &= {\rm Pr}({P_ {\rm A}}{t _{\rm A}} + \sigma _{\rm w}^2 > \Gamma ) \nonumber\\
			&= \left\{ \begin{array}{l}
				1,\quad\thinspace\thinspace \quad\quad\quad\quad\quad\quad\quad\quad\quad\thinspace\thinspace\thinspace\thinspace \thinspace \Gamma<\sigma _{\rm w}^2,\\
				1 - \frac{{\Gamma  - \sigma _{\rm w}^2 - {P_{\rm A,\min }}{t _{\rm A}}}}{{({P_{\rm A,\max }} - {P_{\rm A,\min }}){t _{\rm A}}}},\sigma _{\rm w}^2 < \Gamma<{\Delta _{\rm A}},\\
				0,\quad\quad\quad\quad\quad\quad\quad\quad\quad\quad\quad\thinspace \Gamma>{\Delta _{\rm A}},
			\end{array} \right.
		\end{align}
		where ${\Delta _{\rm A}}={P_{\rm A,\max }}{t _{\rm A}}+\sigma _{\rm w}^2$, ${t _{\rm A}}={\left| {\mathbf{h}_{\rm w}^H{{\boldsymbol x}_{\rm A}}} \right|^2}$, ${t _{{\mathbf{w}}_1}}={\left| {\mathbf{h}_{\rm w}^H{{\mathbf{w}}_{1}}} \right|^2}$ ${\lambda _{{{\mathbf{w}}_1}}}={\bar {\mathbf{w}}^H}\bar {\mathbf{w}}$ and $\bar {\mathbf{w}}=\sqrt{{{l}_{\rm w}}}{{\mathbf{\Omega} _{\rm w}^{\frac{1}{2}}}}{\mathbf{w}}_1$.
		\begin{proof}
			Please refer to Appendix A.	 
		\end{proof}

		\begin{figure*}[hb] 
			\centering 
			\hrulefill 
			\vspace*{8pt} 
			\begin{align}\label{23107}
				{P_{\rm MD}} &= \Pr ({P_{\rm A }}{t _{\rm A}} + {t _{{{\mathbf{w}}_1}}} + \sigma _{\rm w}^2 < \Gamma )\nonumber\\
				&= \left\{ \begin{array}{l}
					0,\quad\quad\quad\quad\quad\quad\quad\quad\quad\quad\quad\quad \quad\quad\quad\quad\quad\quad\quad\quad\quad\thinspace\thinspace\thinspace\thinspace\ \Gamma<\sigma _{\rm w}^2, \\
					\frac{{\Gamma  - \sigma _{\rm w}^2 - {P_{\rm A,\min }}{t _{\rm A}}}+{P_{{\rm A},\min }}{t_{\rm A}}{e^{ - \frac{{\Gamma  - \sigma _{\rm w}^2}}{{{\lambda _{{w_1}}}}}}} + {\lambda _{{{\mathbf{w}}_1}}}({e^{ - \frac{{\Gamma  - \sigma _{\rm w}^2}}{{{\lambda _{{{\mathbf{w}}_1}}}}}}} - 1)}{{({P_{\rm A,\max }} - {P_{\rm A,\min }}){t _{\rm A}}}},\sigma _{\rm w}^2 < \Gamma<{\Delta _{\rm A}},\\
					1-\frac{{ {\lambda _{{{\mathbf{w}}_1}}}{e^{ - \frac{{\Gamma  - \sigma _{\rm w}^2}}{{{\lambda _{{{\mathbf{w}}_1}}}}}}} }}{{({P_{\rm A,\max }} - {P_{\rm A,\min }}){t _{\rm A}}}}({e^{\frac{{{{t _{\rm A}}}{P_{\rm A,\max }}}}{{{\lambda _{{\mathbf{w}_1}}}}}}} - {e^{\frac{{{{t _{\rm A}}}{P_{\rm A,\min }}}}{{{\lambda _{{\mathbf{w}_1}}}}}}}),\quad\quad\quad\thinspace\thinspace\thinspace\thinspace\Gamma  > {\Delta _{\rm A}},
				\end{array} \right.
			\end{align}
		\end{figure*}
		
		As per  (\ref{23106}) and  (\ref{23107}), the DEP at Willie is given by
		
		\begin{align}
			\xi  = \left\{ \begin{array}{l}
				1,\quad\quad\quad\quad\quad\quad\quad\quad\quad\quad\quad\quad\quad\quad\quad \thinspace\thinspace\ \Gamma<\sigma _{\rm w}^2,\\
				1 + \frac{{ {\lambda _{{{\mathbf{w}}_1}}}({e^{ - \frac{{\Gamma  - \sigma _{\rm w}^2}}{{{\lambda _{{{\mathbf{w}}_1}}}}}}} - 1)}}{p_\alpha }+   p_\beta{e^{ - \frac{{\Gamma  - \sigma _{\rm w}^2}}{{{\lambda _{{{\mathbf{w}}_1}}}}}}}                ,\sigma _{\rm w}^2 < \Gamma<{\Delta _{\rm A}},\\
				1-\frac{{ {\lambda _{{{\mathbf{w}}_1}}}{e^{ - \frac{{\Gamma  - \sigma _{\rm w}^2}}{{{\lambda _{{{\mathbf{w}}_1}}}}}}} }}{p_\alpha }({e^{\frac{{{{t _{\rm A}}}{P_{\rm A,\max }}}}{{{\lambda _{{\mathbf{w}_1}}}}}}} - {e^{\frac{{{{t _{\rm A}}}{P_{\rm A,\min }}}}{{{\lambda _{{\mathbf{w}_1}}}}}}}),\Gamma  > {\Delta _{\rm A}},
			\end{array} \right.
		\end{align}
		where ${p_\alpha }={{({P_{\rm A,\max }} - {P_{\rm A,\min }}){t _{\rm A}}}}$ and ${p_\beta }=\frac{{P_{\rm A,\min }}}{{P_{\rm A,\max }} - {P_{\rm A,\min }}}$. We can derive that when $\sigma _{\rm w}^2 < \Gamma<{\Delta _{\rm A}}$, the DEP at Willie is a decreasing function with respect to (w.r.t.) $\Gamma$. However, when $\Gamma  > {\Delta _{\rm A}}$, the DEP at Willie is a increasing function w.r.t. $\Gamma$. Hence the optimal $\Gamma $, denoted by ${\Gamma ^ * }$, is given by ${\Gamma ^ * }={P_{\rm A,\max }}{t _{\rm A}}+\sigma _{\rm w}^2$. Thus the optimal ${\xi ^ * }= 	1 + \frac{{ {\lambda _{{{\mathbf{w}}_1}}}({e^{ - \frac{{P_{\rm A,\max }}{t _{\rm A}}}{{{\lambda _{{{\mathbf{w}}_1}}}}}}} - 1)}}{p_\alpha }+   p_\beta{e^{ - \frac{{P_{\rm A,\max }}{t _{\rm A}}}{{{\lambda _{{{\mathbf{w}}_1}}}}}}}$. Similar to Appendix A, we can prove that ${t _{\rm A}}\sim {\rm exp}({\lambda _{\rm A}})$, where ${\lambda _{\rm A}}={\mathbf{w}_{\rm A}^H}{\mathbf{w}_{\rm A}}$ and $ {\mathbf{w}_{\rm A}}=\sqrt{{{l}_{\rm w}}}{{\mathbf{\Omega} _{\rm w}^{\frac{1}{2}}}}{\boldsymbol x}_{\rm A}$. By averaging ${\xi ^ * }$ over ${t _{\rm A}}$, we can get the
		the average result of the minimum DEP as
		\begin{equation}
			\begin{aligned}
				{\bar \xi }^ *&={\mathbb{E}_{{t_{\rm A}}}}({{\bar \xi }^ * })\\
				&=\int_0^{ + \infty } \left( 1 + \frac{{ {\lambda _{{{\mathbf{w}}_1}}}({e^{ - \frac{{P_{\rm A,\max }}{t _{\rm A}}}{{{\lambda _{{{\mathbf{w}}_1}}}}}}} - 1)}}{p_\alpha }
				\right. \\&
				\left.	+   p_\beta{e^{ - \frac{{P_{\rm A,\max }}{t _{\rm A}}}{{{\lambda _{{{\mathbf{w}}_1}}}}}}} \right) \frac{1}{{{\lambda _{\rm A}}}}{e^{ - \frac{{{t_{\rm A}}}}{{{\lambda _{\rm A}}}}}}{d}{{{t_{\rm A}}}}
				\\&=1+\nu\int_0^{ + \infty }\frac{{{e^{ - (\frac{{{P_{{\rm A},\max }}}}{{{\lambda _{{{\mathbf{w}}_1}}}}} + \frac{1}{{{\lambda _{\rm A}}}}){t_{\rm A}}}} - {e^{ - \frac{{{t_{\rm A}}}}{{{\lambda _{\rm A}}}}}}}}{{{t_{\rm A}}}}d{t_{\rm A}}
				\\&+{p_\beta}\int_0^{ + \infty }\frac{1}{{{\lambda _{\rm A}}}}{e^{ - (\frac{{{P_{{\rm A},\max }}}}{{{\lambda _{{{\mathbf{w}}_1}}}}} + \frac{1}{{{\lambda _{\rm A}}}}){t_{\rm A}}}}d{t_{\rm A}}
				\\&=1+\nu \ln \mu+{p_\beta}\mu,
			\end{aligned}
		\end{equation}
		
		where $\nu=\frac{{\lambda _{{{\mathbf{w}}_1}}}}{{({P_{\rm A,\max }} - {P_{\rm A,\min }}){\lambda _{\rm A}}}} $ and $\mu  = \frac{{{{\lambda _{{{\mathbf{w}}_1}}}}}}{{{{P_{{\rm A},\max }}}{\lambda _{\rm A}}+{{{\lambda _{{{\mathbf{w}}_1}}}}}}}$. 
		To further explore the derived closed-form expression of ${\bar \xi }^ *$, we denote $\pi=\frac{{{P_{\rm A,\max }}}}{{{P_{\rm A,\min }}}}$ and $\tau= \frac{{{\lambda _{\rm A}}{P_{\rm A,\max }}}}{{{\lambda _{{\mathbf{w}_1}}}}}$, thus we can rewrite ${\bar \xi }^ *$ as
		\begin{align}\label{statistical}
			{\bar \xi }^ *(\tau)=1-\left( {\frac{\pi }{{\pi  - 1}}\frac{{\ln (\tau  + 1)}}{\tau }}-\frac{1 }{{\pi  - 1}}\frac{1}{{\tau  + 1}} \right).
		\end{align}
		
		The first-order derivative of ${\bar \xi }^ *(\tau)$ w.r.t. $\tau$ is given by 
		\begin{align}\label{monotonicity}
			\frac{{d{{\bar \xi }^ * }(\tau )}}{{d\tau }}&= - \frac{\pi }{{(\pi  - 1){{(1 + \tau )}^2}{\tau ^2}}}( \tau  - (1 + \tau )\ln (1 + \tau )   \nonumber\\&
			+(1 + \frac{1}{\pi }){\tau ^2} - \tau (1 + \tau )\ln (1 + \tau ) ).
		\end{align}
		Since when $\tau > 0$, $\tau  - (1 + \tau )\ln (1 + \tau ) $ and $(1 + \frac{1}{\pi }){\tau ^2} - \tau (1 + \tau )\ln (1 + \tau )<0$ always holds. Therefore we can derive that $\frac{{d{{\bar \xi }^ * }(\tau )}}{{d\tau }}> 0$. Recall that $\tau= \frac{{{\lambda _{\rm A}}{P_{\rm A,\max }}}}{{{\lambda _{{\mathbf{w}_1}}}}}$, we observe that ${\bar \xi }^ *$ is an increasing function w.r.t. ${P_{\rm A,\max }}$, which indicates that increasing ${P_{\rm A,\max }}$ can degrade the wiretap performance of Willie. Note that in (\ref{monotonicity})  $\pi$ is a constant w.r.t. different ${P_{\rm A,\max }}$ and ${P_{\rm A,\min }}$, which has no influence on the monotonicity of $	{\bar \xi }^ *(\tau)$. Moreover, it can be derived that when ${{P_{\rm A,\max }}  \to +\infty }$, ${\bar \xi }^ *=1$. Despite the potential trade-off between the performance metrics of sensing and covert communications, by appropriately setting the value of $\tau$, both the requirement of sensing and covert communications can be satisfied simultaneously.
		
		\subsection{Robust Beamforming Optimization Design under Statistical WCSI Errors}	
		Considering the statistical WCSI, we delve into the robust beamforming optimization design. Consistent with the objective proposed under imperfect WCSI, to minimize the weighted sum of the sensing beampattern MSE and cross correlation in $ {{\cal H}_1}$, the covert communications
		signal and the DFAN signal are jointly optimized subject to the covert communications requirements under statistical WCSI.

		According to the covert communications performance analysis, the expression of DEP’s average result is derived as ${\bar \xi }^ *(\tau)$ in \eqref{statistical}. Moreover, the following discussion of its monotonicity indicates that ${\bar \xi }^ *$ is an increasing function w.r.t. $\tau$. Considering the general covertness constraint ${\bar{\xi}^ * }\ge1-\epsilon$, firstly we define $f(\tau)={\frac{\pi }{{\pi  - 1}}\frac{{\ln (\tau  + 1)}}{\tau }}-\frac{1 }{{\pi  - 1}}\frac{1}{{\tau  + 1}} $ for ease of expression. Thus the covertness constraint can be transformed to $	f(\tau) \le \epsilon$ and further rewritten as $	\tau \ge \tau _\epsilon$, where $f(\tau _\epsilon)=\epsilon$. Recall that $\tau= \frac{{{\lambda _{\rm A}}{P_{\rm A,\max }}}}{{{\lambda _{{\mathbf{w}_1}}}}}$, where ${\lambda _{{{\mathbf{w}}_1}}}={\bar {\mathbf{w}}^H}\bar {\mathbf{w}}={{l}_{\rm w}}{\mathbf{w}}_1^H\mathbf{\Omega} _{\rm w}{\mathbf{w}}_1$ and 
		${\lambda _{\rm A}}={\mathbf{w}_{\rm A}^H}{\mathbf{w}_{\rm A}}={{l}_{\rm w}}{\boldsymbol x}_{\rm A}^H\mathbf{\Omega} _{\rm w}{\boldsymbol x}_{\rm A}$. Thus, the covertness constraint can be given as
		\begin{align}\label{cs}
			{\boldsymbol x}_{\rm A}^H\mathbf{\Omega} _{\rm w}{\boldsymbol x}_{\rm A}{P_{\rm A,\max }} \ge \tau _\epsilon{\mathbf{w}}_1^H\mathbf{\Omega} _{\rm w}{\mathbf{w}}_1.
		\end{align}

		Due to constraint (\ref{cs}), the increasing of transmit power is confined, which cannot always increase the covert communication rate. The covertness constraint (\ref{cs}) can be further transformed as
		\begin{align}\label{cs2}
			{\rm Tr}({{{\mathbf{\Omega} _{\rm w}}\mathbf{T}} {{} \mathord{\left/
						{\vphantom {{} {}}} \right.
						\kern-\nulldelimiterspace} {}} {{P_{\rm A}}}}) \ge \tau _\epsilon {\rm Tr}({\mathbf{\Omega} _{\rm w}}{\mathbf{W}_1})/{P_{\rm A,\max }}.
		\end{align}

		After applying the DC relaxation method, the covert communication rate constrained beampattern optimization problem is formulated as 
		\begin{subequations}\label{p2}
			\begin{align}\text {(P3)}:
				&\mathop {\rm min}\limits_{\eta_1,\mathbf{W}_1,\mathbf{T}}     	L(\eta_1,\mathbf{W}_1,\mathbf{T}),\label{p2:sub1}
				\\ &s.t.\thinspace\thinspace\thinspace\thinspace(\ref{cs2}),\label{p2:sub2} \\
				&\quad\quad\thinspace{\mathbf{h}_{\rm b}^H{\mathbf{W}_1}\mathbf{h}_{\rm b}}\ge{(2^{{R_{\min }} }- 1)}{{\mathbf{h}_{\rm b}^H{ {\mathbf{T}}} {\mathbf{h}_{\rm b}} + \sigma _{\rm b}^2}},
				\\&\quad\quad\thinspace{\rm Tr}(\mathbf{W}_1)\le {P_{\rm t}}-{P_{\rm A,\max }},
				\\ &\quad\quad(\ref{12}),
			\end{align}	
		\end{subequations}
		where \eqref{p2:sub2} denotes the covertness constraint in the statistical WCSI scenario. Note that in the covertness constraint \eqref{p2:sub2}, $f(\tau _\epsilon)=\epsilon$ should be firstly satisfied to get $\tau _\epsilon$ according to the closed-form expressions of the average minimum DEP derived in section \uppercase\expandafter{\romannumeral3}. Problem $\text {(P3)}$ is a convex
		QSDP problem, which can be optimally tackled utilizing CVX.

		\section{Feasibility-Checking based DC Algorithm Design}
		In this section, a feasibility-checking based DC algorithm design is proposed to solve the optimization problems formulated in the last section, which ensures the initial value of $\mathbf{W}_1$, denoted by $\mathbf{W}_{0}$, in feasible region of the problem and guarantee the convergence of the algorithm.

		As per the covertness constraints (21) and (44) derived in the imperfect WCSI scenario and the statistical scenario, respectively, we can see that the covertness of the covert ISAC system is influenced by the value of $P_{\rm A}$ in both cases. However, due to the fact that $P_{\rm A}$ is the instant power of the DFAN, $P_{\rm A}$ cannot be optimized due to its randomness to create uncertainty at Willie. In the formulated optimization problems, due to the DC relaxation method being adopted to tackle the rank-one constraint, the initial value of $\mathbf{W}_1$ should be first set before the iterations. Thus it is utmost to choose a proper initial value of $\mathbf{W}_1$. To elaborate, the constraints (21) and (44) are recast respectively as 	
		\begin{align}\label{1722}
			P_{\rm A,\min}\le \frac{{\epsilon ({P_{\rm A,\max }} - {P_{\rm A,\min }})	\mathbf{h}_{\rm w}^H{\mathbf{T}}\mathbf{h}_{\rm w}}}{{	\mathbf{h}_{\rm w}^H{\mathbf{W}_1}\mathbf{h}_{\rm w}}},
		\end{align}
		
		\begin{align}\label{cs22}
			P_{\rm A,\min}\le \frac{{{P_{\rm A,\max }}{\rm Tr}({{\mathbf{\Omega} _{\rm w}}\mathbf{T}})}}{{\tau _\epsilon {\rm Tr}({\mathbf{\Omega} _{\rm w}}{\mathbf{W}_1})}}.
		\end{align}

		Likewise, the covert rate constraint can be reformulated as 
		\begin{align}\label{cr1}
			P_{\rm A,\min}\le  \frac{{{\mathbf{h}_{\rm b}^H{\mathbf{W}_1}\mathbf{h}_{\rm b}}-\sigma _{\rm b}^2}}{{(2^{{R_{\min }} }- 1){{\mathbf{h}_{\rm b}^H{ {\boldsymbol x_{\rm A}}\boldsymbol x_{\rm A}^H} {\mathbf{h}_{\rm b}} }}}}.
		\end{align}

		Thus the feasibility-checking problem for the imperfect WCSI case and the statistical WCSI case can be formulated respectively given as \text {(P4)} and \text {(P5)} as follows.
		\begin{subequations}\label{191111}
			\begin{align}\text {(P4)}:
				&\rm{Find} \thinspace\mathbf{W}_1
				\\ &s.t.\thinspace\thinspace\thinspace\thinspace(26),(\ref{1722}),(\ref{cr1}),\\
				&\quad\quad\thinspace{\rm Tr}(\mathbf{W}_1)\le {P_{\rm t}}-{P_{\rm A,\max }},
			\end{align}	
		\end{subequations}
		\begin{subequations}\label{p212}
			\begin{align}\text {(P5)}:
				&\rm{Find} \thinspace\mathbf{W}_1
				\\ &s.t.\thinspace\thinspace\thinspace\thinspace(26),(\ref{cs22}),(\ref{cr1}), \\
				&\quad\quad\thinspace{\rm Tr}(\mathbf{W}_1)\le {P_{\rm t}}-{P_{\rm A,\max }},
			\end{align}	
		\end{subequations}
		where in the imperfect WCSI case, i.e., \text {(P4)}, the problem can be further reformulated in the bounded and Gaussian WCSI scenarios, which are omitted here for brevity. After checking the feasibility of the initial value of $\mathbf{W}_1$, the covert communications
		signal covariance matrix $\mathbf{W}_1$ and the DFAN signal covariance matrix $\mathbf{T}$ are jointly optimized such that the weighted sum of the sensing beampattern MSE and cross correlation is minimized in $ {{\cal H}_1}$,
		subject to the covert communications requirements. Based on the procedures above, the overall algorithm for optimal transmit beampattern design is summarized in Algorithm 1, where $L_{\rm th}$ denotes the predefined threshold of the reduction of the objective function.
		\begin{algorithm}[!h]
			\caption{Feasibility-Checking Based DC Algorithm }
			\begin{algorithmic}[1]
				\STATE Initialize $\mathbf{T}(n)$ and $\mathbf{W}_0$ by solving 
				the feasibility-checking problem \text {(P4)} or \text {(P5)} corresponding to the imperfect or the statistical WCSI case, respectively;
				\REPEAT
				
				\REPEAT
				\STATE{For given iteration parameters, solve problem \text {(P1.2)} or \text {(P2)} or \text {(P3)} } corresponding to different WCSI cases;
				\STATE{Set $n=n+1$};
				\STATE{	Update the iteration parameters:  ${\mathbf{w}_l} (n)$ is replaced by the leading eigenvector of ${\mathbf{W}_l} (n-1)$;}
				
				\UNTIL{$L(\eta_1(n),\mathbf{W}_1(n),\mathbf{T}(n))-L(\eta_1(n-1),\mathbf{W}_1(n-1),\mathbf{T}(n-1))\le L_{\rm th}$};
				\STATE Set ${\varrho _{\rm w_1}} (n)=k{\varrho _{\rm w_1}} (n-1), k>0$; 
				\UNTIL{${\varrho _{\rm w_1}} (n) \le {\varrho _{\rm th}}$}.
				
			\end{algorithmic}
		\end{algorithm}

		\begin{figure} [t!]
			\centering
			\includegraphics[width=3.2in]{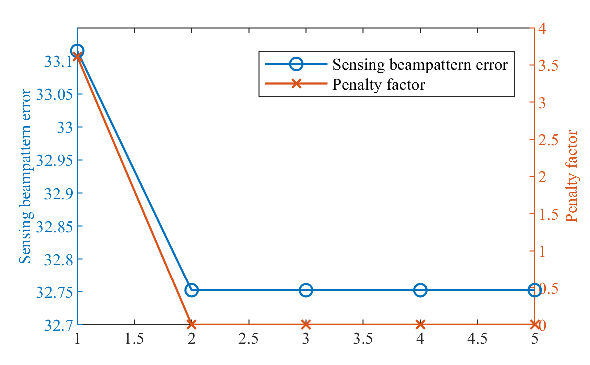}
			\caption{Convergence of the proposed feasibility-checking based DC algorithm
				when $R_{\rm min}$ = 8 bps/Hz.
			}
			
		\end{figure}

		\begin{figure}[t!]
			\centering
			\subfigure[Covert rate versus $P_{\rm{b,max}}$ under bounded WCSI.]{\label{SOP_asym}
				\includegraphics[width=3.2in]{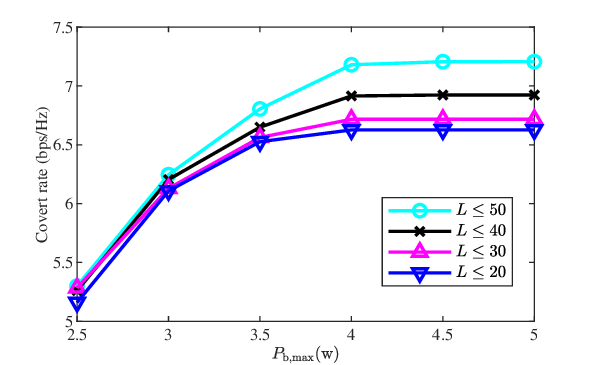}}
			\subfigure[Covert rate versus $P_{\rm{b,max}}$ under Gaussian WCSI.]{\label{SOP_pair}
				\includegraphics[width=3.2in]{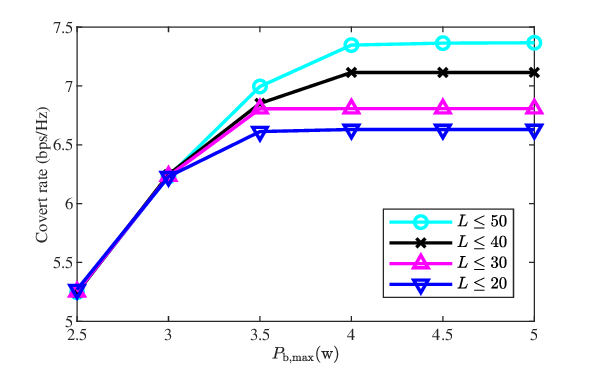}}
			\subfigure[Covert rate versus $P_{\rm{b,max}}$ under statistical WCSI.]{\label{SOP_pair_zone}
				\includegraphics[width=3.2in]{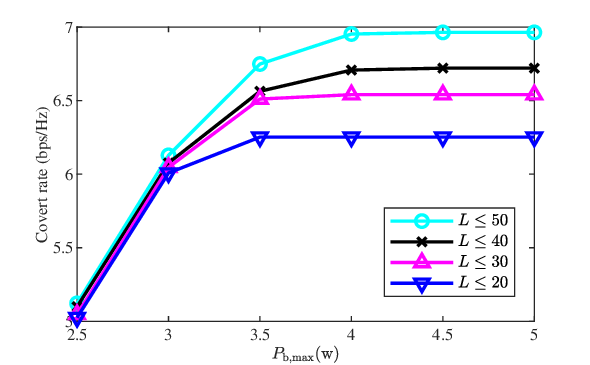}}
			
			\caption{Covert rate versus $P_{\rm{b,max}}$ under bounded, Gaussian and statistical WCSI.}\label{11223123}

		\end{figure}

		\begin{figure} [t!]
			\centering
			\includegraphics[width=3.2in]{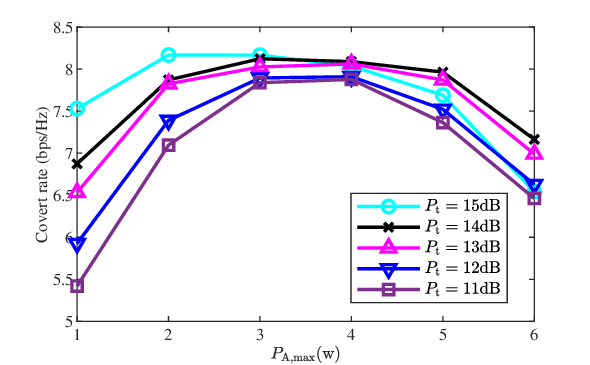}
			\caption{Covert rate versus $P_{\rm{A,max}}$ under statistical WCSI with $ 	L(\eta_1,\mathbf{W}_1,\mathbf{T})\le 200$ and $P_{\rm{A,max}}=\frac{{x}}{{7}}10^{P_{\rm{t}}/10}\rm{ w}.$
			}
			
		\end{figure}
		\begin{figure} [t!]
			\centering
			\includegraphics[width=3.2in]{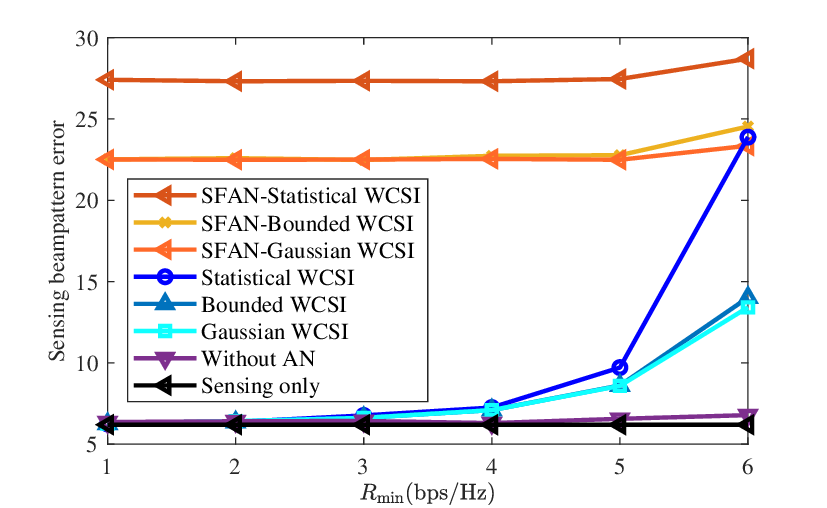}
			\caption{Sensing beampattern error versus  $R_{\rm min}$ with $P_{\rm{b,max}}\le5\rm{ w}.$ 
			}
			
		\end{figure}

		\begin{figure*}[t!]
			\centering
			\subfigure[Transmit beampattern of the ideal interference cancellation scheme.]{\label{SOP_asym}
				\includegraphics[width= 2.3in, height=1.5in]{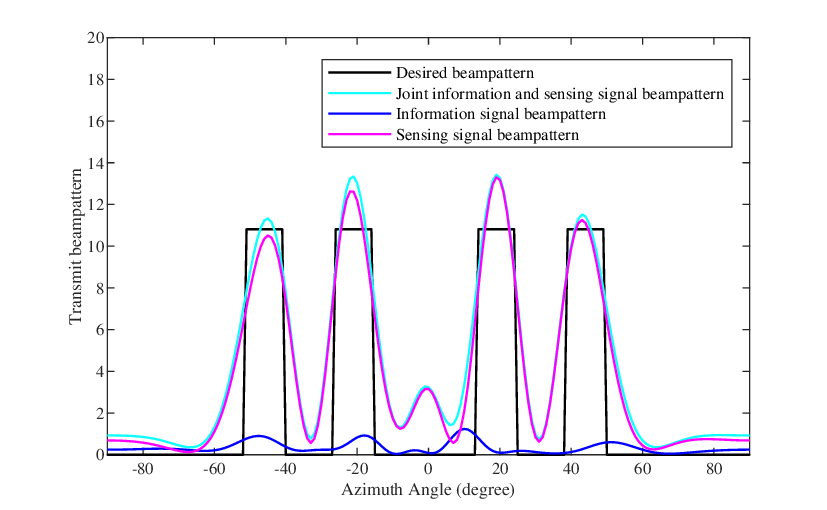}}
			\subfigure[Transmit beampattern of the proposed DFAN signal scheme.]{\label{SOP_pair}
				\includegraphics[width= 2.3in, height=1.5in]{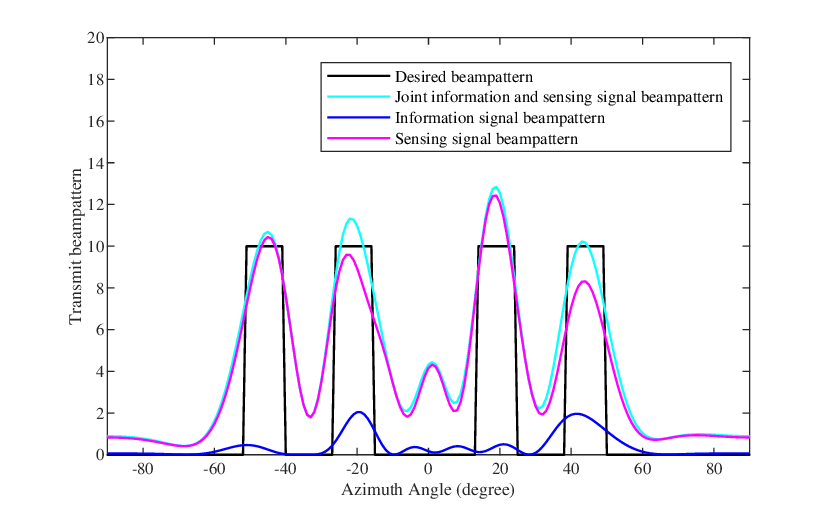}}
			\subfigure[Transmit beampattern of the dedicated sensing signal scheme.]{\label{SOP_pair_zone}
				\includegraphics[width= 2.3in, height=1.5in]{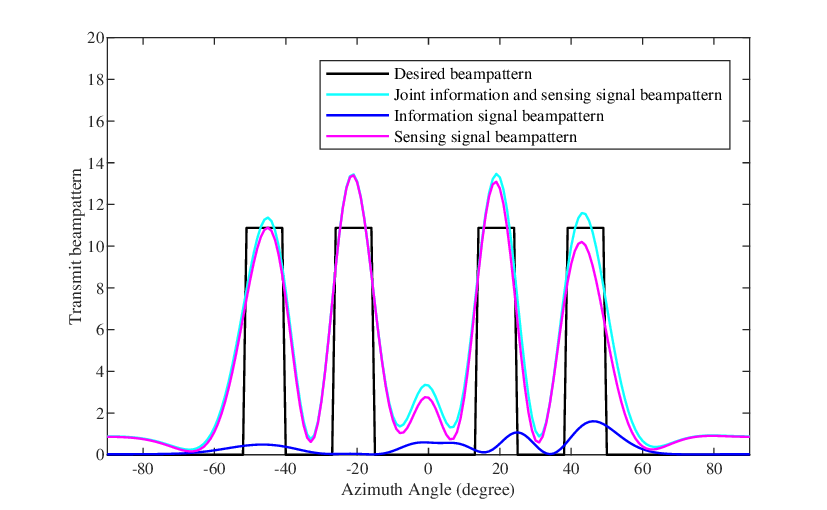}}
			
			\caption{The transmit beampattern corresponding to the scenarios considered in Fig. \ref{p5_noma_beampattern_2}.}\label{SOP_single_antenna}

		\end{figure*}

		\begin{figure} [t!]
			\centering
			\includegraphics[width=3.2in]{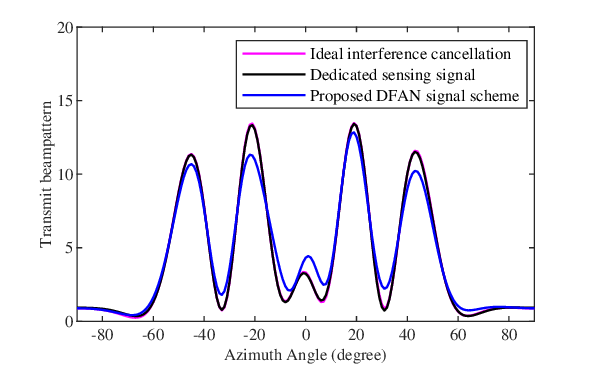}
			\caption{Transmit beampattern under bounded WCSI with $R_{\rm min}$ = 5 bps/Hz and $P_{\rm{b,max}}\le5\rm{ w}.$ 
			}\label{p5_noma_beampattern_2}
			
		\end{figure}

		\section{Simulation Results}
		In this section, we provide numerical results to evaluate the
		performance of our proposed DFAN design for covert ISAC systems. Unless specified otherwise, the simulation parameters
		are set as follows. Alice is equipped with $N$ = 10 transmit antennas and the normalized spacing
		between two adjacent antennas is set as $\frac{d}{\lambda } = 0.5$. The large-scale path loss is modeled as ${{l}_{ {xy} }} =  {{\zeta _0}{{({{{d_0}} \mathord{\left/
							{\vphantom {{{d_0}} d}} \right.
							\kern-\nulldelimiterspace} d})}^\alpha }} $ for all channels, where ${\zeta _0}$ is the path loss
		at the reference distance ${d_0} = 1$ m, $\alpha$ is the path-loss exponent,
		and $d$ is the link distance. The large-scale path loss of the Alice-Bob link and the Alice-Willie link are denoted as ${{l}_{ {\rm b} }}$ and ${{l}_{ {\rm w} }}$, respectively. For small-scale fading, the Alice-Willie channel and the Alice-Bob channel are assumed to be Rayleigh fading, i.e., ${\mathbf{h}}_{\rm w}\sim {\cal C}{\cal N}({\bold 0},\mathbf{I})$ and ${\mathbf{h}}_{\rm b}\sim {\cal C}{\cal N}({\bold 0},\mathbf{I})$, respectively. The distance between Alice and $M$ targets/Bob is set as 50 m.  The path loss at the
		reference distance of 1 meter, i.e., ${\zeta _0}$, is set as ${\zeta _0}=-30{\rm dB}$. The path-loss exponents of
		the Alice-Bob channel and the Alice-Willie channel are set as $\alpha_{\rm b}=\alpha_{\rm e}=2.5$. The noise power
		at Bob and Willie is set as $\sigma _{\rm{b}}^2=\sigma _{\rm{e}}^2=-80{\rm dBm}$. The outage probability is set as ${\rho _c}$=0.05. For the Bounded and Gaussian WCSI error model, we set $\varepsilon_{\rm w}=0.1\left| {{\hat {\mathbf{h}}_{\rm w}}} \right| $ and ${\boldsymbol{\gamma }_{\rm w}}=0.01{\left\| {\hat {\mathbf{h}}_{\rm w}}  \right\|^2}{\mathbf{I}_N}/N$, respectively. Besides concealing the signal transmission to Bob located in the direction of ${0^\circ }$, Alice also
		senses $M=4$ targets in the directions of ${-45^\circ }$, ${-20^\circ }$, ${20^\circ }$ and ${45^\circ }$, thus ${\hat \theta _m}$ is given as ${\hat \theta _m}=\{{{-45^\circ },{-20^\circ }, {20^\circ }, {45^\circ \}}}$. We choose $S = 180$ angles for $\{\theta _s\}$ and the desired beam width $\Delta \theta$ is set as ${10^\circ }$. The covertness constant
		is set as $\epsilon= 0.1$, $P_{\rm{A,max}}$ and $P_{\rm{A,min}}$ are set as 10 w and 1 w, respectively. Moreover, the instant power of the DFAN is set as 5 w.

		The convergence performance of the proposed feasibility-checking based DC algorithm over a specific channel realization when $R_{\rm min}$ = 8 bps/Hz is shown in Fig. 2. We can observe that the sensing beampattern error can quickly converge to a stable value. Moreover, as the number of iterations increases, the penalty factor is gradually reduced to almost zero, which guarantees that the derived $\mathbf{W}_1$ is nearly rank-one. The fast convergence also shows the efficiency
		of the proposed algorithm, which is because the feasibility problem is firstly solved to derive a initial value of $\mathbf{W}_1$ in the feasible region of the problem. Thus the convergence of the algorithm is guaranteed and accelerated.

		Figures 3(a)-(c) show the covert rate versus $P_{\rm{b,max}}$ under bounded, Gaussian and statistical WCSI, in which the acceptable maximum sensing beampattern error $	L(\eta_1,\mathbf{W}_1,\mathbf{T})$ is set as
		20; 30; 40 and 50, where $P_{\rm{b,max}}$ denotes the power allocated for covert communications. It is observed that when $P_{\rm{b,max}}$ increases, the covert rate first increases while finally converging to a certain value. The reasons are that as $P_{\rm{b,max}}$ increases, more transmit power can be allocated to information
		signals towards Bob, leading to the increment of the covert rate. However, the covertness constraint becomes more stringent with the increase of $P_{\rm{b,max}}$, and the transmit power allocated to information signals towards Bob is confined to guarantee covert communications.
		It is also observed that smaller acceptable maximum sensing beampattern error leads to a larger covert rate, which is due to that higher sensing performance requirements limit the available design DoF of the information signal and more power should be allocated for sensing. Moreover, it is observed that in the statistical WCSI scenario, the rate of convergence is relatively slow compared to that in the bounded and Gaussian WCSI scenarios. This is because compared to the other two cases, the covertness constraint is more stringent in the statistical scenario, resulting in the decrease of the available design DoF of the DFAN. Thus more power should be allocated for sensing to compensate for the sensing performance deterioration of the DFAN. Comparing the converged value of the three considered scenarios under the same sensing performance requirement, it can be obtained that the gap between the covert rate achieved in the statistical scenario and that achieved in the other two cases is narrow. This shows the robustness of our proposed DFAN covert ISAC system design that the joint sensing and covert communications performance in the statistical scenario is comparable to the other two cases.

		Fig. 4 depicts the covert rate versus $P_{\rm{A,max}}$ under statistical WCSI, where $ x$ denotes the x-axis value and $P_{\rm{A}}$ is set as $P_{\rm{A}}=( P_{\rm{A,max}}-1)\rm{w}$. Note that due to the inequality holds that $P_{\rm{b,max}}\le P_{\rm{t}}-P_{\rm{a,max}}$, Fig. 4 also depicts the effects of variations of $P_{\rm{A,max}}$ and $P_{\rm{b,max}}$ on the considered system. It is observed that when $P_{\rm{A,max}}$ increases the covert rate increases at first and then decreases. This is because the trade-off between covertness and covert rate induced by the variation of power allocation between $P_{\rm{A,max}}$ and the information signal. As $P_{\rm{A,max}}$ increases, the covertness constraint is relaxed, and thus more power can be allocated to the information signal to improve the covert rate. However, the increment of $P_{\rm{A,max}}$ cannot always improve the covert rate, which is due to the fact that less power is allocated to the information signal when the covertness constraint is further relaxed. Furthermore, it can also be observed that generally increasing the total transmit power can significantly increase the covert rate. Nonetheless, when the total transmit power is sufficiently large, the covertness constraint is stringent. This causes serious deficiency of the power allocated to the information signal and thus degrades the covert rate.

		Fig. 5 plots the sensing beampattern error versus  $R_{\rm min}$. To verify the effectiveness of the proposed DFAN covert ISAC system design, the following benchmark schemes are considered for comparison:

		\textbullet $ \textbf{Sensing only}$: Both the DFAN and the information signal are exploited to minimize the sensing beampattern errors without covert rate requirement.

		\textbullet  $ \textbf{Single functional AN (SFAN)}$: The AN is only harnessed to confuse Willie to achieve covert communications without sensing function. Thus in $ {{\cal H}_0}$, the system has no sensing function. In $ {{\cal H}_1}$, the information signal is fully exploited to achieve covert communications and sense targets simultaneously. Moreover, the three considered WCSI error models are all considered, which are denoted as SFAN-Bounded WCSI, SFAN-Gaussian WCSI, and SFAN-Statistical WCSI, respectively.

		\textbullet $ \textbf{Without AN}$: Due to the lack of AN, covert communications cannot be achieved in this case, the information signal is fully exploited to sense targets, i.e., minimizing the sensing beampattern errors, while satisfying the minimum non-covert communication rate requirement.
		
		\begin{figure*}[hb!] 
			\centering 
			\hrulefill 
			\vspace*{8pt} 
			\begin{align}\label{appendix}
				{P_{\rm MD}} &= \Pr ({P_{\rm A }}{t _{\rm A}} + {t _{{{\mathbf{w}}_1}}} + \sigma _{\rm w}^2 < \Gamma )\nonumber\\
				&= \left\{ \begin{array}{l}
					0, \quad\quad\quad\quad\quad\quad\quad\quad\quad\thinspace\thinspace\thinspace\thinspace\ \Gamma<\sigma _{\rm w}^2, \\
					\int\limits_0^{\Gamma  - \sigma _w^2} {\frac{{(\Gamma  - \sigma _w^2 - {t_{{w_1}}})/{t_A} - {P_{A,\min }}}}{{{P_{A,\max }} - {P_{A,\min }}}}} \frac{1}{{{\lambda _{{w_1}}}}}{{\mathop{\rm e}\nolimits} ^{ - \frac{{{t_{{w_1}}}}}{{{\lambda _{{w_1}}}}}}}d{t_{{w_1}}},\\
					1 - \int\limits_{{P_{A,\min }}}^{{P_{A,\max }}} {{{\mathop{\rm e}\nolimits} ^{ - \frac{{\Gamma  - \sigma _w^2 - {P_A}{t_A}}}{{{\lambda _{{w_1}}}}}}}\frac{1}{{{P_{A,\max }} - {P_{A,\min }}}}d} {P_A},
				\end{array} \right.\nonumber\\
				&= \left\{ \begin{array}{l}
					0,\quad\quad\quad\quad\quad\quad\quad\quad\quad\quad\quad\quad \quad\quad\quad\quad\quad\quad\quad\quad\quad\thinspace\thinspace\thinspace\thinspace\ \Gamma<\sigma _{\rm w}^2, \\
					\frac{{\Gamma  - \sigma _{\rm w}^2 - {P_{\rm A,\min }}{t _{\rm A}}}+{P_{{\rm A},\min }}{t_{\rm A}}{e^{ - \frac{{\Gamma  - \sigma _{\rm w}^2}}{{{\lambda _{{w_1}}}}}}} + {\lambda _{{{\mathbf{w}}_1}}}({e^{ - \frac{{\Gamma  - \sigma _{\rm w}^2}}{{{\lambda _{{{\mathbf{w}}_1}}}}}}} - 1)}{{({P_{\rm A,\max }} - {P_{\rm A,\min }}){t _{\rm A}}}},\sigma _{\rm w}^2 < \Gamma<{\Delta _{\rm A}},\\
					1-\frac{{ {\lambda _{{{\mathbf{w}}_1}}}{e^{ - \frac{{\Gamma  - \sigma _{\rm w}^2}}{{{\lambda _{{{\mathbf{w}}_1}}}}}}} }}{{({P_{\rm A,\max }} - {P_{\rm A,\min }}){t _{\rm A}}}}({e^{\frac{{{{t _{\rm A}}}{P_{\rm A,\max }}}}{{{\lambda _{{\mathbf{w}_1}}}}}}} - {e^{\frac{{{{t _{\rm A}}}{P_{\rm A,\min }}}}{{{\lambda _{{\mathbf{w}_1}}}}}}}),\quad\quad\quad\thinspace\thinspace\thinspace\thinspace\Gamma  > {\Delta _{\rm A}},
				\end{array} \right.
			\end{align}
		\end{figure*}	
		
		From Fig. 5 we can observe that the proposed DFAN design achieves significantly lower sensing beampattern errors than that of the single functional AN design benchmark schemes within the regime of $R_{\rm min}$. Moreover, when $R_{\rm min}$ is relatively small, the sensing beampattern errors achieved by the proposed DFAN design approach that of the Without AN benchmark and the sensing only benchmark under both imperfect and statistical WCSI. The reasons are that for the single functional AN design, the information signal is fully exploited to achieve covert communications and sense targets simultaneously, which highly restricts the available design DoF of the information signal that even a small covert rate requirement leads to great sensing beampattern errors, especially in the statistical WCSI case. However, the proposed DFAN design harnesses the AN to assist in sensing, which shares the sensing function with the information signal, thus can unleash its potential to achieve a distinct higher covert rate than the single functional AN design. Moreover, in $ {{\cal H}_0}$, the single functional AN design has no sensing function, which further limits its application scenarios. Due to the fact that no covertness is required in the Without AN design, the sensing beampattern errors are much close to the sensing only benchmark and are obviously lower than that achieved by the schemes with covertness requirements, especially when $R_{\rm min}$ is relatively large. Furthermore, in contrast to the Without AN benchmark and the sensing only benchmark, the proposed DFAN design achieves comparable sensing performance even in the statistical WCSI scenario when $R_{\rm min}$ is relatively small. Also, among the three cases considered in the proposed DFAN design, the sensing performance of the statistical WCSI case approaches that of the imperfect WCSI cases when $R_{\rm min}$ is small. This shows the robustness of our proposed DFAN design and is consistent with the analyses in Fig. 3. Combining the observations and analyses in Fig. 3 and Fig. 5, two guidelines are given below: 1) To achieve covert communications in an ISAC system, DFAN is preferred rather than single-functional AN. 2) When the covert rate or the sensing performance requirement is low, it is preferable to adopt the statistical WCSI hypothesis to improve the robustness of the covert ISAC design, while achieving a comparable performance as shown in Fig.3 and Fig.5.

		Figs. 6 and 7 depict the transmit beampattern under bounded WCSI and the detailed analyses corresponding to the three considered scenarios. Two benchmark schemes are considered for comparison:
		
		\textbullet $ \textbf{Dedicated sensing signal}$: In this scheme the dedicated sensing signal is utilized to facilitate sensing performance, where the sensing interference induced by the dedicated sensing signal can be canceled by successive interference cancellation (SIC) technologies. However, in this scenario covert communications cannot be achieved. To elaborate, due to the fact that dedicated sensing signal is exploited for multibeam transmission, the covariance matrix ${\mathbf{T}}$ is assumed to be of a general rank with
		$0 \le {\rm rank}({\mathbf{T}})=n \le N$. By exploiting the eigenvalue decomposition of ${\mathbf{T}}$, the dedicated sensing signal ${{\boldsymbol x}_{\rm D}}(k)$ can be decomposed into $n$
		linearly and statistically independent sensing beams, i.e., $	{\mathbf{T}} = \sum\limits_{i = 1}^n {{\lambda _i}{\mathbf{x}_i}{\mathbf{x}_i^H}}  = \sum\limits_{i = 1}^n {{\mathbf{w}_{{\rm a},i}}\mathbf{w}_{ {\rm a},i}^H} ,
		$
		where ${\lambda _i} \in\mathbb{R} $ is the eigenvalue and ${\mathbf{x}}_{ i}\in {{\mathbb{C}}^{N\times 1}}$ is the corresponding eigenvector. The vector ${\mathbf{w}_{a,i}} = \sqrt {{\lambda _i}} {\mathbf{x}_i}$ is the transmit beamformer for ${{\boldsymbol x}_{\rm D}}(k)$. The dedicated sensing signal in $ {{\cal H}_1}$ can be rewritten as $	{{\boldsymbol x}_{\rm D}}(k)= \mathop \sum \limits_{i \in {\cal D}} {\mathbf{w}_{a,i}}{x_{{\rm v},i}(k)} +{\bar {\boldsymbol x}_{\rm D}}(k)  ,$
		where ${\cal D} = \{ 1,....,D\} $ ($1 \le D \le n$) denotes the set of virtual communication signals and the symbols ${\{ {x_{{\rm v},i}(k)}\} _{i \in {\cal D}}}$ are independent as well as with zero mean
		and unit power. It is assumed that ${\{ {x_{{\rm v},i}(k)}\} _{i \in {\cal D}}}$ are independent with ${\bar {\boldsymbol x}_{\rm D}}(k) $. Hence, the covariance matrix ${\bar {\mathbf{T}}}$ of the rest of the dedicated sensing signal ${\bar {\boldsymbol x}_{\rm D}}(k) $ is given by $	{\bar {\mathbf{T}}} = \mathbb{E}\left[ {\bar {\boldsymbol x}_{\rm D}}(k){{\bar {\boldsymbol x}{_{\rm D}}}^H}(k)  \right] \nonumber =\sum\limits_{i = D + 1}^n {{\mathbf{w}_{a,i}}{\mathbf{w}_{a,i}}^H} .$ The signal received at the Bob is given by
		\begin{align}
			{R_{\rm b}} = {\log _2}(1 + \frac{{{{\left| {\mathbf{h}_{\rm b}^H{\mathbf{w}_1}} \right|}^2}}}{{\mathbf{h}_{\rm b}^H{ \bar{{\mathbf{T}}}} {\mathbf{h}_{\rm b}} + \sigma _{\rm b}^2}}).
		\end{align}

		We also assume that only one beam of the
		dedicated sensing signal is embedded with information, i.e., $D= 1$.

		\textbullet $ \textbf{Ideal interference cancellation}$: Bob can perfectly eliminate the sensing interference, which serves as a performance upper bound of the considered scenario.

		From Fig.7, we can observe that compared with the other benchmark schemes, the proposed DFAN signal scheme achieves a sensing beampattern with comparable performance, with peaks in the target directions and small power leakage in the undesired directions. Moreover, the sensing beampattern errors of the proposed DFAN signal scheme, the dedicated sensing signal scheme, and the ideal interference cancellation scheme are given as 8.6519, 7.3543, and 7.2862, respectively. This is because the dedicated sensing signal can be fully leveraged to facilitate sensing and communication performance without covertness constraints, leading to a smaller sensing beampattern error compared to the other two schemes. Moreover, the relatively narrow gap between the proposed DFAN signal scheme and the other two schemes further demonstrates the effectiveness of the proposed DFAN signal design. The sensing performance difference between the proposed DFAN signal scheme and the dedicated sensing signal scheme shows that the DoF of the dedicated sensing signal can be fully exploited to facilitate communication performance. However, the randomness in the DFAN signal to achieve covert communications will inevitably degrade the covert communications performance, which unveils the trade-off between covertness and covert communications performance.

		We dig deep into the three considered scenarios in Fig.6, it can be observed that in the three scenarios, the information signals are all confined in the desired sensing directions, which is due to the sensing beampattern errors, i.e., the gap between the desired beampattern and the joint information and sensing signal beampattern are minimized. Moreover, it can be observed in the three cases that the sensing signal beampattern achieves a non-zero value in the direction of 0 degree. The reasons are that the sensing signal is also constrained by the covertness of the covert ISAC system. A non-zero value in the direction of 0 degree is utilized to conceal the signal transmission of Bob. It can also be observed that in the proposed DFAN signal scheme, the information signal beampattern achieves significantly peaks in the desired sensing directions compared to the other two schemes. The reasons are that in the proposed DFAN signal scheme, in the information signal beampattern, to overcome the interference caused by the DFAN, the information signal peaks in several sensing directions to satisfy the covertness requirement. However, due the interference is partially canceled by SIC technologies in the dedicated sensing signal scheme and it is assumed that the interference is completely canceled in the ideal interference cancellation scheme, the peaks in the information signal beampattern is relatively small compared to that in the proposed DFAN signal scheme. The angular resolution depicted in Fig.6 is consistent with the analyses of Fig.7. To elaborate, although the DFAN signal is introduced to achieve covert communications, it can also degrade the sensing performance, which will induce relatively lower angular resolution.

		\section{Discussions}
		In the section, the case that Willie can acquire the instantaneous CSI of ${\mathbf{h}}_{\rm w}$ is further investigated. Considering the fact that the instantaneous CSI of channel ${{\mathbf{h}}_{\rm w}}$ is not available at Alice, we adopt the minimum DEP averaging over ${t _{{\mathbf{w}}_1}}$ as the covertness metric of the considered covert ISAC system. Note that the minimum DEP $\xi$ is already derived in (20) and ${t _{{\mathbf{w}}_1}}={\left| {\mathbf{h}_{\rm w}^H{{\mathbf{w}}_{1}}} \right|^2}$ is an exponential random variable proved in Appendix A, in the following, we provide a closed-form expression for the minimum average DEP.
		\begin{align}
			{\bar \xi _2}^*& = \int\limits_0^{ + \infty } {(1 + \frac{{ - {\rho _2}}}{{({P_{{\rm{A}},\max }} - {P_{{\rm{A}},\min }}){\rho _1}}})} {f_{t _{{\mathbf{w}}_1}}}(x)dx \nonumber \\
			&\mathop  = \limits^{(a)}  1 - \frac{{{P_b}}}{{{P_{{\rm{A}},\max }} - {P_{{\rm{A}},\min }}}},	
		\end{align}
		where (a) holds by using the CDF of ${t _{{\mathbf{w}}_1}}$, i.e., 
		${F_{t _{{\mathbf{w}}_1}}}(x) = 1 - e^{ - {\lambda _{\mathbf{w}_1}}x}$. Let $	{\bar \xi _2}^*\ge1-\epsilon$, we can derive the covertness constraint as 
		\begin{align}
			{P_b} \le\epsilon ({P_{{\rm{A}},\max }} - {P_{{\rm{A}},\min }}).
		\end{align}

		Compared to the covertness constraint in the case that Willie cannot acquire the instantaneous CSI of ${\mathbf{h}}_{\rm w}$, in this scenario, the covertness constraint is more stringent.
		
		\section{Conclusion}
		In this paper, we studied covert communications in an ISAC system, where the DFAN was harnessed to confuse Willie and sense the targets simultaneously. The robust design considered not only the imperfect WCSI but also the statistical WCSI, where the worst-case scenario was also considered that Willie can adaptively adjust the detection threshold to achieve the best detection performance, and the minimum DEP at Willie was derived in closed form. Moreover, in the statistical WCSI case, the closed-form expressions of the average minimum DEP are derived. The formulated sensing beampattern error minimization problems were tackled by a feasibility-checking based DC algorithm utilizing S-procedure, Bernstein-type inequality, and the DC relaxation method. Simulation results validated the feasibility of the proposed scheme. The results also revealed that: 1) To achieve covert communications in an ISAC system, DFAN is preferred rather than single-functional AN. 2) When the covert rate or the sensing performance requirement is low, it is preferable to adopt the statistical WCSI hypothesis to improve the robustness of the covert ISAC design, while achieving comparable performance.

		\section*{Appendix A: Derivation of Detection Error Probability}
		According to the decision rule $
		{\varsigma}\underset{{{D}_{0}}}{\overset{{{D}_{1}}}{\mathop{\gtrless }}}\,\Gamma
		$ and the uniform
		distribution of variable $P_{\rm A}$, the false alarm probability of Willie can be derived as
		\begin{align}
			{P_{\rm FA}} &= {\rm Pr}({P_ {\rm A}}{t _{\rm A}} + \sigma _{\rm w}^2 > \Gamma ) \nonumber\\
			&= \left\{ \begin{array}{l}
				1,\quad\thinspace\thinspace \quad\quad\quad\quad\quad\quad\quad\quad\quad\thinspace\thinspace\thinspace\thinspace \thinspace \Gamma<\sigma _{\rm w}^2,\\
				1 - \frac{{\Gamma  - \sigma _{\rm w}^2 - {P_{\rm A,\min }}{t _{\rm A}}}}{{({P_{\rm A,\max }} - {P_{\rm A,\min }}){t _{\rm A}}}},\sigma _{\rm w}^2 < \Gamma<{\Delta _{\rm A}},\\
				0,\quad\quad\quad\quad\quad\quad\quad\quad\quad\quad\quad\thinspace \Gamma>{\Delta _{\rm A}}.
			\end{array} \right.
		\end{align}
		To further analyze the miss detection probability of Willie, due to only the statistical CSI of ${\mathbf{h}}_{\rm w}$ being available at Willie, we define ${t _{{\mathbf{w}}_1}}={\left| {\mathbf{h}_{\rm w}^H{{\mathbf{w}}_{1}}} \right|^2}$ and ${\lambda _{{{\mathbf{w}}_1}}}={\bar {\mathbf{w}}^H}\bar {\mathbf{w}}$, where $\bar {\mathbf{w}}=\sqrt{{{l}_{\rm w}}}{{\mathbf{\Omega} _{\rm w}^{\frac{1}{2}}}}{\mathbf{w}}_1$. It can be obtained that ${t _{{\mathbf{w}}_1}}$ is an exponential random variable whose PDF is ${f_{{t_{{w_1}}}}}(x) = \frac{1}{{{\lambda _{{w_1}}}}}{{\mathop{\rm e}\nolimits} ^{ - \frac{{{t_{{w_1}}}}}{{{\lambda _{{w_1}}}}}}}$. Thus the false alarm probability of Willie can be derived as \eqref{appendix} at the bottom of the previous page.

				\bibliographystyle{IEEEtran}
				\bibliography{citation}
				\begin{IEEEbiography}[{\includegraphics[width=1in,height=1.25in,clip,keepaspectratio]{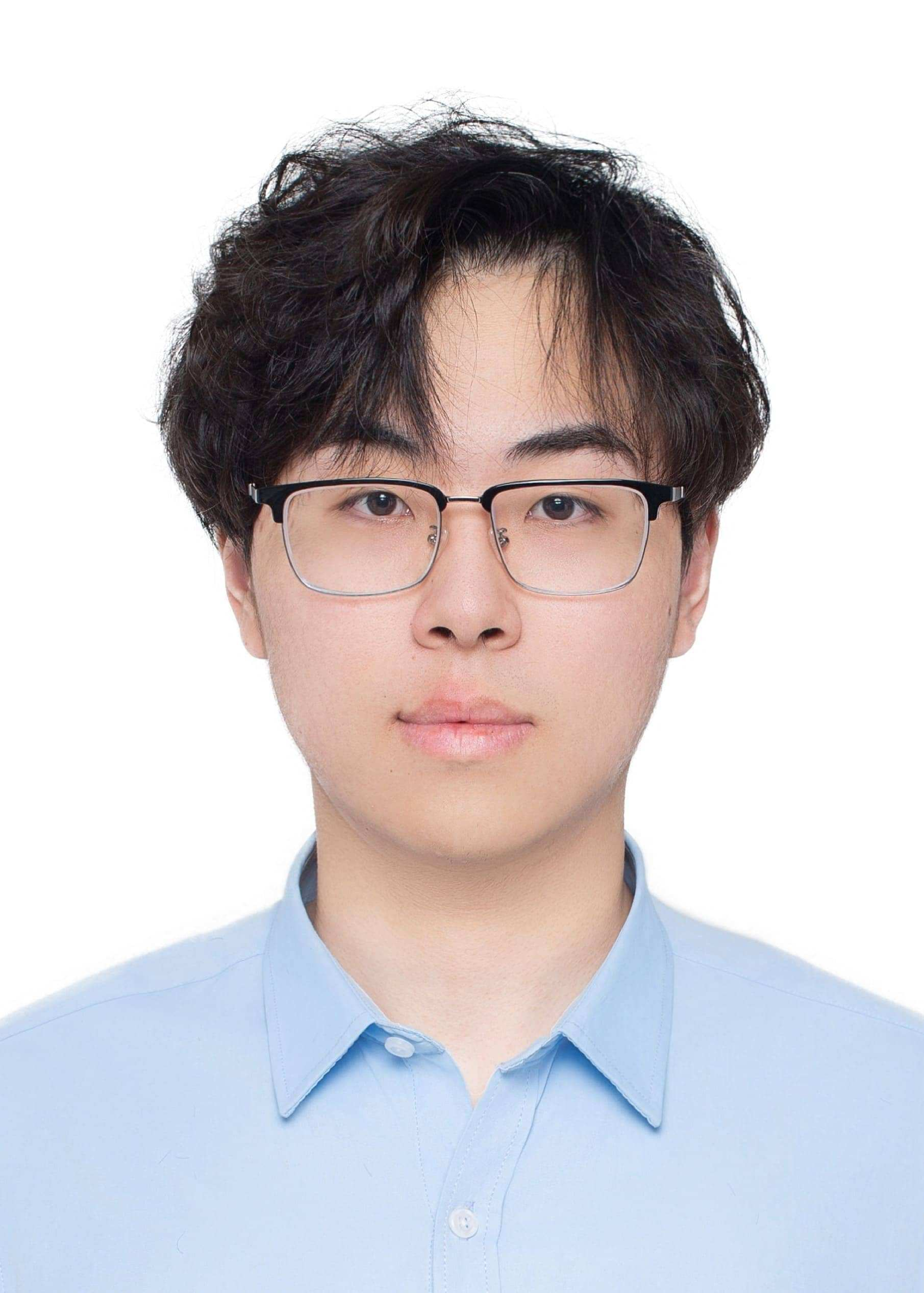}}]{Runzhe Tang}
					received the B.E. degree from Nanjing University of Posts and Telecommunications (NJUPT), Nanjing, China, in 2023. He is currently pursuing the M.S. degree with the State Key Laboratory of Integrated Services Networks, Xidian University, Xi'an, China. His current research interests include wireless physical-layer security, covert communications, and integrated sensing and communications.
				\end{IEEEbiography}
			
			\begin{IEEEbiography}[{\includegraphics[width=1in,height=1.25in,clip,keepaspectratio]{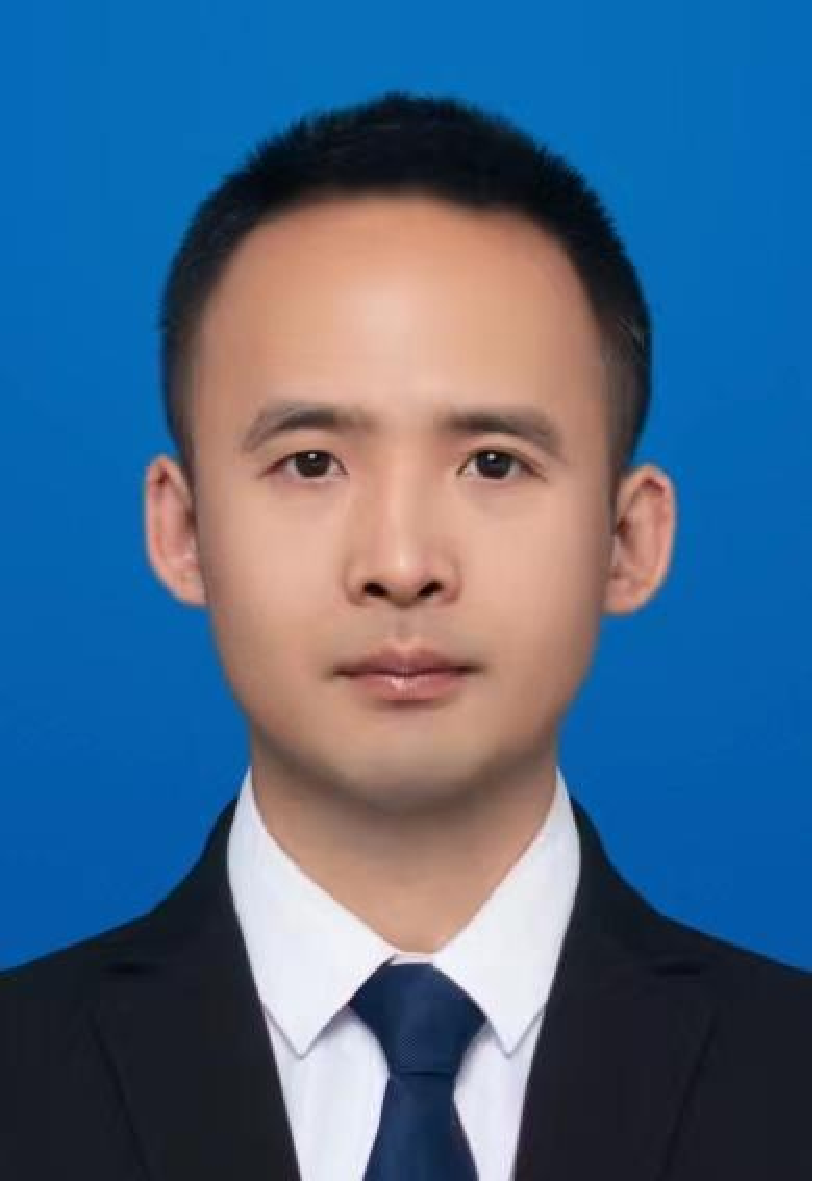}}]{Long Yang}
				(Senior Member, IEEE) received the B.E. and Ph.D. degrees from Xidian University, Xi'an, China, in 2010 and 2015, respectively. Since 2015, he has been a Faculty Member with Xidian University, where he is currently an Associate Professor with the State Key Laboratory of Integrated Services Networks.
				From 2017 to 2019, he was also a Post-doctoral Fellow with the Department of Electrical and Computer Engineering, University of Alberta, Canada. His current research interests include non-orthogonal multiple access, cooperative communications and wireless physical-layer security. He is now serving as a Lead Guest Editor for \textsc{IEEE Internet of Things Journal} and an Associate Editor for \textsc{IET Communications}.
			\end{IEEEbiography}
		
		\begin{IEEEbiography}[{\includegraphics[width=1in,height=1.25in,clip,keepaspectratio]{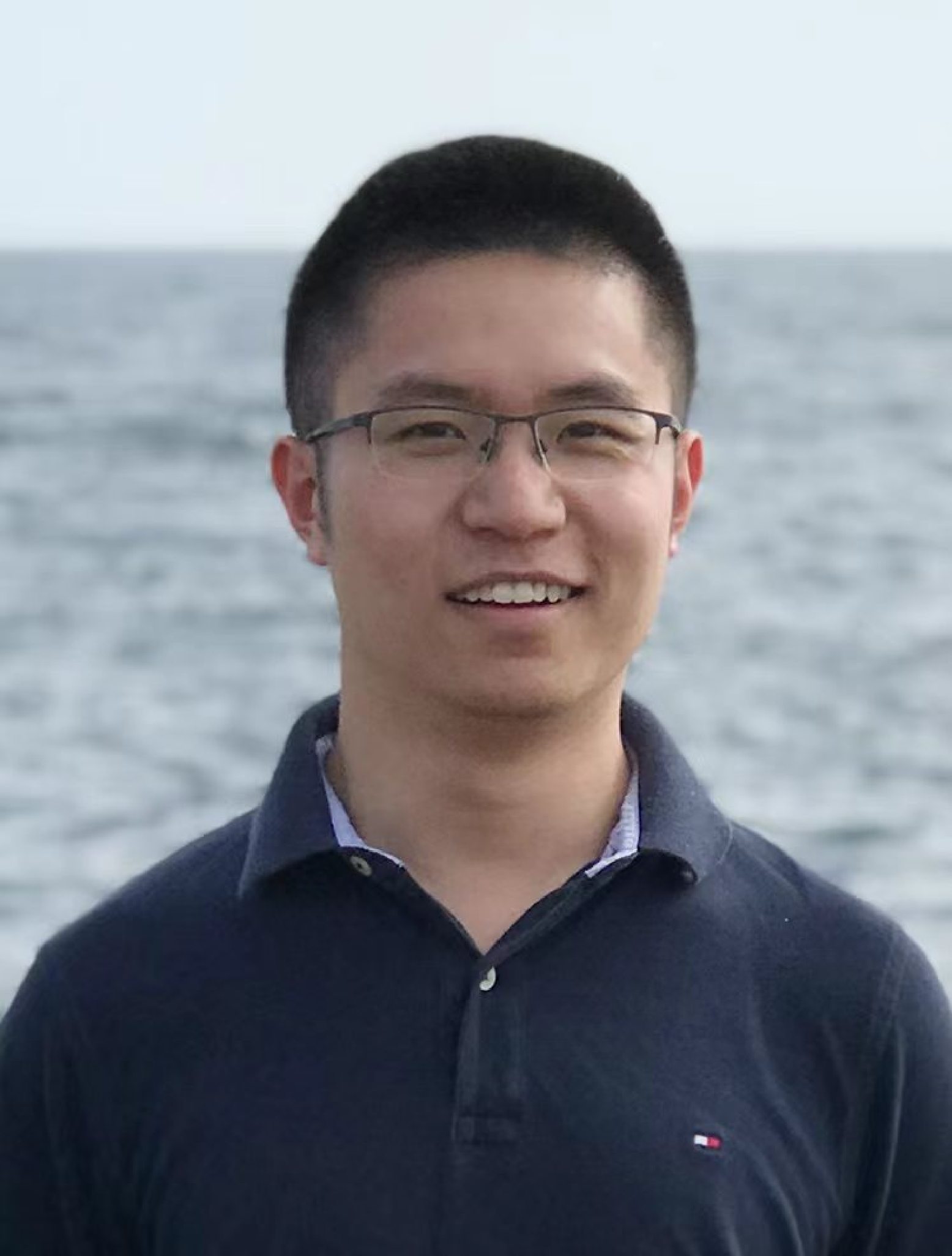}}]{Lu Lv}
			(Member, IEEE) received the Ph.D. degree from Xidian University, Xi’an, China, in 2018.
			From 2016 to 2018, he was an Academic Visitor with Lancaster University, Lancaster, U.K., and the University of Alberta, Edmonton, AB, Canada. In 2019, he was a Postdoctoral Fellow with Dalhousie University, Halifax, NS, Canada. He is currently an Associate Professor with Xidian University. His research interests include non-orthogonal multiple access, reconfigurable intelligent surfaces, physical layer security, and covert communication.

			Dr. Lv was a recipient of the Outstanding Ph.D. Thesis Award of Shannxi Province in 2020, IEEE ICCC Best Paper Award in 2021, and the Exemplary Reviewer Certificate for \textsc{IEEE Transactions on Communications} from 2018 to 2020 and \textsc{IEEE Communications
			Letters} in 2022. He was listed as a World’s Top 2$\%$ Scientist by Stanford University in 2022 and 2023. He serves as an Associate Editor for \textsc{IEEE Internet of Things Journal}, \textsc{IEEE Open Journal of The Communications Society}, and Frontiers in Computer Science.
		\end{IEEEbiography}
		
		\begin{IEEEbiography}[{\includegraphics[width=1in,height=1.25in,clip,keepaspectratio]{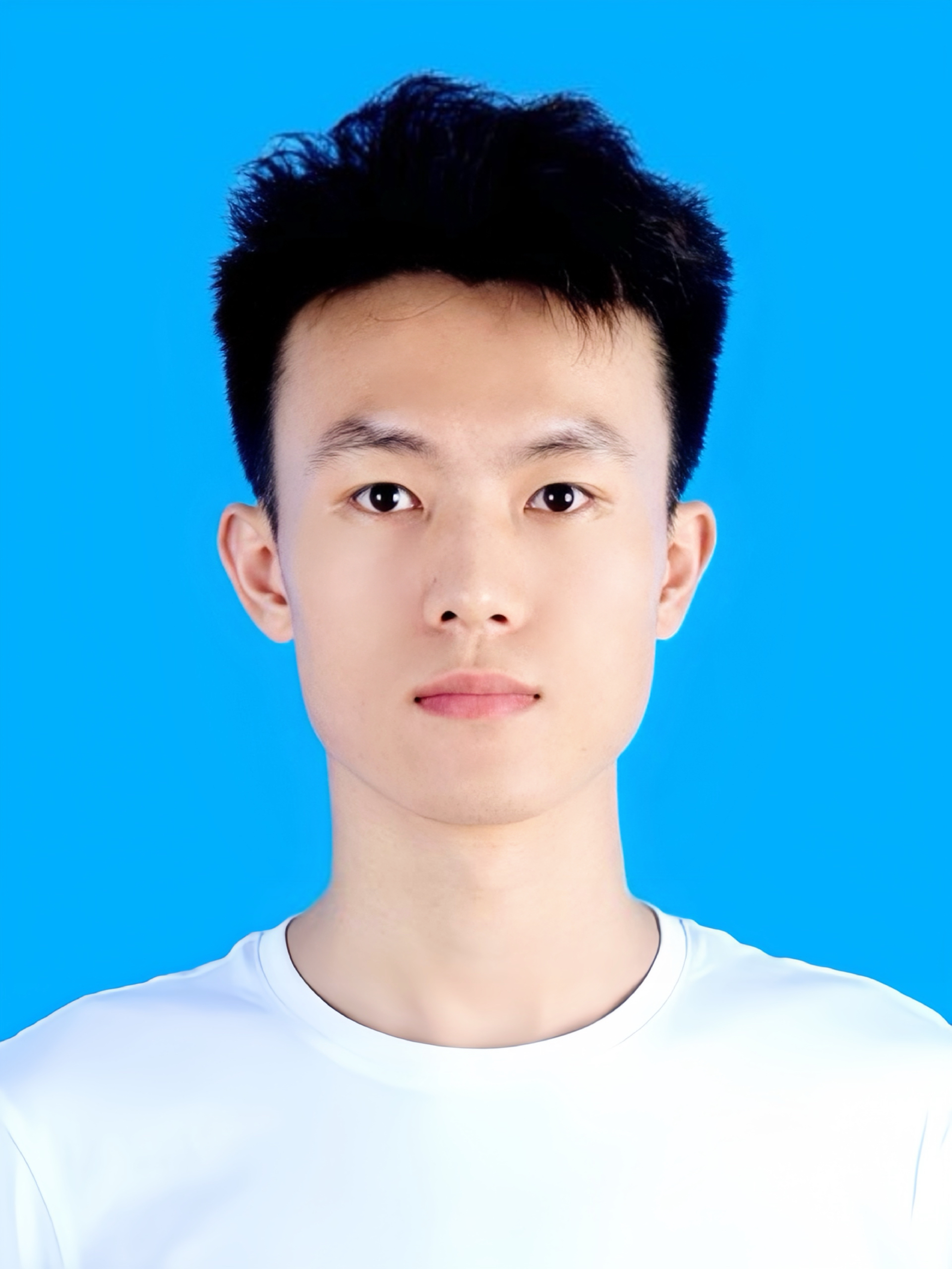}}]{Zheng Zhang}
			(Graduate Student Member, IEEE)
			received the B.S. degree from Xidian University, Xi'an, China, in 2020, where he is currently pursuing the Ph.D. degree. From 2022, he was a joint Ph.D. Student with Queen Mary University of London U.K., sponsored by the China Scholarship Council. His research interests include reconfigurable intelligent surface, physical-layer security, non-orthogonal multiple access, and integrated sensing and communications. He is the recipient of the Excellent Graduate Student Award in IEEE Ucom 2023 and IEEE Ucom 2024.
		\end{IEEEbiography} 
		
			\begin{IEEEbiography}[{\includegraphics[width=1in,height=1.25in,clip,keepaspectratio]{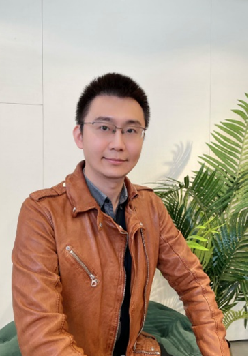}}]{Yuanwei Liu} (S'13-M'16-SM'19-F’24, \url{https://www.eee.hku.hk/~yuanwei/}) has been a (tenured) full Professor in Department of Electrical and Electronic Engineering (EEE) at The University of Hong Kong (HKU) since September, 2024. Prior to that, he was a Senior Lecturer (Associate Professor) (2021-2024) and a Lecturer (Assistant Professor) (2017- 2021) at Queen Mary University of London (QMUL), London, U.K, and a Postdoctoral Research Fellow (2016-2017) at King's College London (KCL), London, U.K. He received the Ph.D. degree from QMUL in 2016.  His research interests include non-orthogonal multiple access, reconfigurable intelligent surface, near field communications, integrated sensing and communications, and machine learning.
				
				Yuanwei Liu is a Fellow of the IEEE, a Fellow of AAIA, a Web of Science Highly Cited Researcher, an IEEE Communication Society Distinguished Lecturer, an IEEE Vehicular Technology Society Distinguished Lecturer, the rapporteur of ETSI Industry Specification Group on Reconfigurable Intelligent Surfaces on work item of “Multi-functional Reconfigurable Intelligent Surfaces (RIS): Modelling, Optimisation, and Operation”, and the UK representative for the URSI Commission C on “Radio communication Systems and Signal Processing”. He was listed as one of 35 Innovators Under 35 China in 2022 by MIT Technology Review. He received IEEE ComSoc Outstanding Young Researcher Award for EMEA in 2020. He received the 2020 IEEE Signal Processing and Computing for Communications (SPCC) Technical Committee Early Achievement Award, IEEE Communication Theory Technical Committee (CTTC) 2021 Early Achievement Award. He received IEEE ComSoc Outstanding Nominee for Best Young Professionals Award in 2021. He is the co-recipient of the 2024 IEEE Communications Society Heinrich Hertz Award, the Best Student Paper Award in IEEE VTC2022-Fall, the Best Paper Award in ISWCS 2022, the 2022 IEEE SPCC-TC Best Paper Award, the 2023 IEEE ICCT Best Paper Award, and the 2023 IEEE ISAP Best Emerging Technologies Paper Award. He serves as the Co-Editor-in-Chief of IEEE ComSoc TC Newsletter, an Area Editor of \textsc{IEEE Transactions on Communications} and \textsc{IEEE Communications Letters}, an Editor of \textsc{IEEE Communications Surveys $\&$ Tutorials}, \textsc{IEEE Transactions on Wireless Communications}, \textsc{IEEE Transactions on Vehicular Technology}, \textsc{IEEE Transactions on Network Science and Engineering}, and \textsc{IEEE Transactions on Cognitive Communications and Networking}. He serves as the (leading) Guest Editor for \textsc{Proceedings of the IEEE on Next Generation Multiple Access}, \textsc{IEEE JSAC on Next Generation Multiple Access}, \textsc{IEEE JSTSP on Intelligent Signal Processing} and \textsc{Learning for Next Generation Multiple Access}, and \textsc{IEEE Network on Next Generation Multiple Access for 6G}. He serves as the Publicity Co-Chair for IEEE VTC 2019-Fall, the Panel Co-Chair for IEEE WCNC 2024, Symposium Co-Chair for several flagship conferences such as IEEE GLOBECOM, ICC and VTC. He serves the academic Chair for the Next Generation Multiple Access Emerging Technology Initiative, vice chair of SPCC and Technical Committee on Cognitive Networks (TCCN).

		\end{IEEEbiography} 
	
	\begin{IEEEbiography}[{\includegraphics[width=1in,height=1.25in,clip,keepaspectratio]{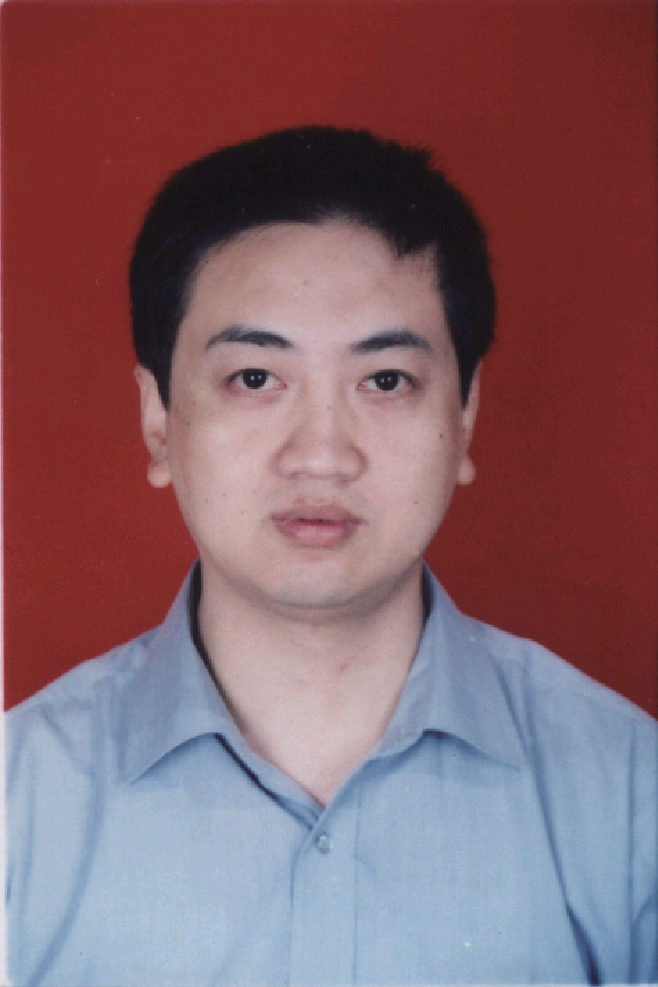}}]{Jian Chen}
		(Member, IEEE) received the B.S. degree from Xi'an Jiaotong University, China, in 1989, the M.S. degree from Xi'an Institute of Optics and Precision Mechanics of Chinese Academy of Sciences in 1992, and the Ph.D. degree in Telecommunications Engineering in Xidian University, China, in 2005. He is a Professor with the State Key Laboratory of Integrated Services Networks (ISN), Xidian University. He was a visitor scholar in the University of Manchester from 2007 to 2008. His research interests include cognitive radio, OFDM, wireless sensor networks and non-orthogonal multiple access.
	\end{IEEEbiography}

			\end{document}